\documentclass[preprint, twocolumn,10pt, twoside, notitlepage]{revtex4-2}%
\usepackage{amsmath}%
\usepackage{amsfonts}%
\usepackage{amssymb}%
\usepackage{graphicx}
\usepackage[caption=false]{subfig}
\graphicspath{{pics/}}
\usepackage{aps_math}
\usepackage{qm}
\usepackage{ifthen}
\def\bibfile{My_Library}% for central bibfile
\ifthenelse{\equal{\bibfile}{}}{%
  \def\myprintbibliography{}%
}{%
  \def\myprintbibliography{%
    \bibliographystyle{myunsrt}%
    \bibliography{\bibfile}%
  }%
}%
\usepackage[%
]{hyperref}
\def\myprintglossary{%
}
\begin{document}
%%%%%%%%%%%%%%%%%%%%%%%%%%%%%%%%%%%%%%%%%%%%%%%%%%%%%%%%%%%%%%%%%%%%%%
%%                         Title page                               %%
%%                                                                  %%
\title{Generation and detection of discrete-variable multipartite entanglement \\with multi-rail encoding in linear optics networks}
\author{Jun-Yi Wu}
\email{junyiwuphysics@gmail.com}
\affiliation{Department of Physics, Tamkang University, 151 Yingzhuan Rd., Tamsui Dist., New Taipei City 25137, Taiwan, ROC}
\begin{abstract}
A linear optics network is a multimode interferometer system, where indistinguishable photon inputs can create nonclassical interference that can not be simulated with classical computers. Such nonclassical interference implies the existence of entanglement among its subsystems, if we divide its modes into different parties. Entanglement in such systems is naturally encoded in multi-rail (multi-mode) quantum registers. For bipartite entanglement, a generation and detection scheme with multi-rail encoding has been theoretically proposed [NJP 19(10):103032, 2017] and experimentally realized [Optica, 7(11):1517, 2020]. In this paper, we will take a step further to establish a theory for the detection of multi-rail-encoded discrete-variable genuine multipartite entanglement (GME) in fixed local-photon-number subspaces of linear optics networks. We also propose a scheme for GME generation with both discrete-variable (single photons) and continuous-variable (squeezed states) light sources. This scheme allows us to reveal the discrete-variable GME in continuous-variable systems. The effect of photon losses is also numerically analyzed for the generation scheme based on continuous-variable inputs.
\end{abstract}
\keywords{Linear optics network, boson sampling, genuine multipartite entanglement, squeezed state}

\maketitle
%\let\newpage\original_newpage
%%                                                                  %%
%%                         Title page                               %%
%%%%%%%%%%%%%%%%%%%%%%%%%%%%%%%%%%%%%%%%%%%%%%%%%%%%%%%%%%%%%%%%%%%%%%

%%%%%%%%%%%%%%%%%%%%%%%%%%%%%%%%%%%%%%%%%%%%%%%%%%%%%%%%%%%%%%%%%%%%%%
%%                         Main text                                %%
%%                                                                  %%
  %%%%%%%%%%%%%%%%%%%%%%%%%%%%%%%%%%%%%%%%%%%%%%%%%%%%%%%%%%%%%%%%%%
  %%    Introduction
\section{Introduction}

%
%\begin{enumerate}
%  \item MltEnt
%    \begin{enumerate}
%      \item in distinguishable register of single particle
%    \end{enumerate}
%  \item MltEnt in LON
%    \begin{enumerate}
%      \item (indistinguishable particles)
%      \item mode as information register
%      \item particle number as DV
%      \item bipartite Ent \cite{WuHofmann2017-BiEntMltMd,WuMurao2020-CmplPropLONs}
%    \end{enumerate}
%  \item it allows single photon and CV sources to generate DV MltEnt
%  \item HW symmetry.
%\end{enumerate}

For quantum information processing in multipartite systems, multipartite entanglement has its advantages over bipartite entanglement, if the local systems have limited sizes \cite{YamasakiEtAlBarbara2018-MltEntOutpfmBiEnt}.
In particular, genuine multipartite entanglement (GME) \cite{AcinBrussEtAlSanpera2001-3QbtCls} plays an important role in measurement-based quantum computation \cite{RaussendorfBriegel2001-MBQC}, quantum algorithms \cite{BrussMacchiavello2011-MEntInQAlgo}, quantum secret sharing \cite{HilleryBuzekBerthiaume1999-QSS, MarkhamSanders2008-GSforQSS, BellHMWRTame2014-ExpGSQSS}, and quantum conference key distribution \cite{DasEtAlHorodecki2021-LmtOnQKD, DasEtAlDowling2018-RbstQNetEntDistr}.
The generation and verification of genuine multipartite entanglement are therefore essential steps for various quantum information application.
Many theories have been proposed for experimental generation and verification of GME in multipartite qudit systems \cite{GuhneEtAl2003-ExpDetEntLocMeas, BourennaneETAL2004-WitnessingMtEnt, TothGuhne2005-EntDetectStabilizer, TothGuhne2005-GMEDetection, SpenglerHuberEtAlHiesmayr2012-EntWitViaMUB, MacconeBrussMacchiavello2015-CmplCrr, HuangEtAlPeruzzo2016-HghDimEntCert, SauerweinAtElKraus2017-MltptCrrMUBs}.

As a quantum computing that has already outperformed its classical counterpart \cite{ZhongEtAlLuPan2020-QCAcvBsnSmpl,ZhongEtAlLuPan2021-PhsProgGBS}, boson sampling can generate classically non-simulatable photon statistics \cite{AaronsonArkhipov2011-CmplxLinOps}.
One would therefore expect a large amount of entanglement in such systems.
However, the role of entanglement in boson sampling computing is still unclear.
The problem is that the indistinguishability of photons is the prerequisite of the quantum advantage of such processing \cite{RenemaEtAlWalmsley2018-PhDstngForBS}.
In linear optics networks (LONs), where boson sampling is implemented, the indistinguishability of photons tangles the concept of entanglement \cite{%First
EckertSchliemannEtAlLewenstein2002-QCorrIndistPtcls,
%Memory register
DowlingDohertyWiseman2006-EntIndistPtcl, KilloranCramerPlenio2014-IdPtclEnt, ChinHuh2019-EntIdPtclCoh1QL,
%CmplSetProp
GhirardiMarinattoWeber2002-Ent, GhirardiMarinatto2004-EntCritIdP,
%Tichy
Tichy2011Thesis, TichyEtAlBuchleitner2013-DtctLvlEnt, TichyMintertBuchleitner2013-LimMEntBsnFmn,
%TAlgbra
ReuschSperlingVogel2015-IdParticleEntWit, GrabowskiKusMarmo2011-EntMltptIndstPcl,
%IT
FrancoCompagno2016-EntIdPtclByITNotion, LourencoDebarbaDuzzioni2019-EntIndstPtcl}, since entanglement should be defined with distinguishable local quantum information registers.
In such a system, path modes are naturally distinguishable quantum information registers. Note that mode entanglement is widely considered in continuous-variable (CV) systems, in which entanglement can be detected with covariance matrices \cite{Reid1989-CVEntWit, DuanEtAlZoller2000-CVEntWit, Simon2000-CVEntWit, HyllusEisert2006-OptEntWitCV}.
However, in boson sampling (e.g. Gaussian boson sampling \cite{HamiltonEtAlJex2017-GBS}), one resolves photon numbers and postselects measurement outputs in fixed-photon-number subspaces, where the concept of continuous-variable entanglement also becomes inadequate.

One solution to this problem is to treat the path modes as distinguishable multi-rail registers, on which the discrete-variable photon occupation number is encoded as the quantum digits. Such discrete-variable multi-rail encoding is represented as multi-mode Fock states $\ket{\boldvec{n}}$,
\begin{equation}
  \ket{\boldvec{n}}: = \ket{n_{0},...,n_{M-1}},
\end{equation}
where $n_{m}$ is the photon number in the $m$-th mode.
The entanglement defined in such $M$-rail encoding was first quantified in \cite{WisemanVaccaro2003-IdPtclEnt}.
Although one can extract conventional multipartite qudit entanglement with one photon in each party from a multipartite multiphoton LON system \cite{KilloranCramerPlenio2014-IdPtclEnt}, the extraction will change the testing states.
A detection method without entanglement extraction was established in \cite{WuHofmann2017-BiEntMltMd, WuMurao2020-CmplPropLONs} for bipartite entanglement in multiphoton $M$-rail-encoded LONs.
Such entanglement was generated and verified in bipartite multiphoton three-rail LON systems in experiments \cite{KiyoharaEtAlTakeuchi2020-VerfEntBiptLONs}.

%
%A conventional multipartite qudit system is multiple distinguishable particles.
%Each of the particles is a well-defined distinguishable quantum information register $\ket{e_{m}}$.
%In LONs, it is equivalent to a single-photon $M$-mode system.
%\begin{equation}
%\label{eq::single_photon_M-rail}
%  \ket{e_{m}} = \ket{0_{0},...,0_{m-1},1_{m},0_{m+1}...,0_{M-1}}.
%\end{equation}
%Each mode is a well-defined path in the outputs or inputs of a LON.
%Such quantum information register is called an $M$-rail qudit.

In this paper, we will address a further question about the generation and evaluation of multi-rail discrete-variable genuine multipartite entanglement in multiphoton LONs, which might have further application for quantum conference key distribution \cite{DasEtAlHorodecki2021-LmtOnQKD}. We will establish a theory that allows us to detect GME in fixed-photon-number subspaces of both discrete-variable and continuous-variable LON systems.

In particular, our target GME exhibits Heisenberg-Weyl (HW) symmetry, which generates special photon statistics under generalized Hadamard transformation \cite{WuMurao2020-CmplPropLONs, DittelEtAlKeil2018-DestrInterfPermSymMPtclSt, DittelEtAlKeil2018-DestrMPtclInterf}.
In Section \ref{sec::criteria}, we will derive a GME criterion employing a GME verifier \cite{Wu2020-AdptQSFEBiEnt} that stabilizes the target Heisenberg-Weyl symmetric states.
This criterion can be evaluated directly in experiments implemented with local generalized Hadamard transformation.

For potential experimental implementation, we propose a generation scheme for GME in Section \ref{sec::generation}. Our generation scheme allows single-photon sources and continuous-variable sources, such as displaced squeezed vacuum. We evaluate and compare the GME generated by single-photon and squeezed sources.
In particular, for the evaluation of GME with CV inputs, we add displacement on the input squeezed vacuums. Such a displacement operation is a local unitary operation, which does not change the overall entanglement of the whole CV system. However, in fixed-photon-number subspaces, one can reveal and observe the diminishing of the GME signature through our GME verifier (witness). This demonstrates the possible application of our method for revealing the transfer of GME between different photon-number subspaces under the conservation of total GME of a CV system.

%We will evaluate the GME generated by single-photon and squeezed sources, respectively.
%
%
%The paper is structured as follows.
%In Section \ref{sec::criteria}, we will derive a GME criterion employing a GME verifier \cite{Wu2020-AdptQSFEBiEnt} that stabilizes the target Heisenberg-Weyl symmetric states. This criterion can be evaluated directly in experiments implemented with local generalized Hadamard transformation.
%In Section \ref{sec::generation}, we will propose a generation scheme for GME. Our generation scheme allows single-photon sources and continuous-variable sources, such as displaced squeezed vacuum. We will evaluate and compare the GME generated by single-photon and squeezed sources, respectively.
%We will then conclude the paper in Section \ref{sec::conclusion}.
  %%%%%%%%%%%%%%%%%%%%%%%%%%%%%%%%%%%%%%%%%%%%%%%%%%%%%%%%%%%%%%%%%%
  %%
\section{Criteria for GME in LONs}
\label{sec::criteria}
%In an $M$-mode linear optics network, an $N$-photon pure state is intrinsically represented as a superposition of Fock states in the 2nd quantization formalism
%\begin{equation}
%  \ket{\psi_{N}}
%  =
%  \sum_{\boldvec{n}: |\boldvec{n}| = N}
%  c_{\boldvec{n}}\ket{\boldvec{n}}.
%\end{equation}
%where $\boldvec{n} = (n_{0},..., n_{M-1})$ is a vector of photon occupation number in each mode indexed by $m\in\{0, ..., M-1\}$, and $\ket{\boldvec{n}}$ are $M$-mode Fock states.

From a practical point of view, a criterion for GME that can be evaluated with local operations and measurements is desirable.
We therefore consider a $P$-partite linear optics network system, in which each local system is an $M$-mode local LON labeled by $i\in\{1, ..., P\}$ and no global interference between two parties is allowed.
Without the global interference among parties, the photon number in each local system is conserved under perfect linear optics operations.
We therefore consider a pure state $\ket{\psi_{N_{1},N_{2},..., N_{P}}}$ with fixed local photon number $N_{i}$ in the $i$-th local LON,
\begin{equation}
  \ket{\psi_{N_{1},N_{2},..., N_{P}}} =
  \sum_{\boldvec{n}_{i}: |\boldvec{n}_{i}| = N_{i}}
  c_{\boldvec{n}_{1}, ..., \boldvec{n}_{P}}\ket{\boldvec{n}_{1}, ..., \boldvec{n}_{P}}.
\end{equation}
Here, the vector $\boldvec{n}_{i} = (n_{0}^{(i)},..., n_{M-1}^{(i)})$ indicates the photon occupation number $n_{m}^{(i)}$ in the $m$-th mode of the $i$-th local system.
In such $P$-partite multiphoton $M$-rail encoding systems, a state $\ket{\phi_{\text{bisep.}}}$ is biseparable, if a bipartition exists such that $\ket{\phi_{\text{bisep.}}}$ is a product state.
A state $\widehat{\rho}$ is biproducible, if it can be decomposed as a mixture of biseparable states $\ket{\phi_{\text{bisep.}}}$.
On the contrary, it is called genuinely multipartite (GM) entangled, if it cannot be decomposed as a mixture of biseparable states $\ket{\phi_{\text{bisep.}}}$
\begin{equation}
  \widehat{\rho}_{\text{GME}} \neq \sum_{p_{\phi},\phi_{\text{bisep.}}} p_{\phi} \projector{\phi_{\text{bisep.}}}.
\end{equation}
%The state $\ket{\psi_{N_{1},...,N_{P}}}$ is biseparable, if there exists a bipartition of local modes, in which $\ket{\psi_{N}}$ can be factorized into a product state.
%A mixed state is biproducible\cite{GuhneTothBriegel2005-MEntInSpinChain}, if it can be decomposed as a mixture of biseparable pure states,
%\begin{equation}
%  \widehat{\rho}_{\text{biprod.}} = \sum_{p_{\phi},\phi_{\text{bisep.}}} p_{\phi} \projector{\phi_{\text{bisep.}}},
%\end{equation}
%On the contrary, a state $\widehat{\rho}$ is mathematically defined as genuinely multipartite (GM) entangled, if it is not biproducible in the 2nd quantization formalism,
%\begin{equation}
%  \widehat{\rho}_{\text{GME}} \neq \sum_{p_{\phi},\phi_{\text{bisep.}}} p_{\phi} \projector{\phi_{\text{bisep.}}}.
%\end{equation}

%If the photon number in each local mode is $N_{i}=1$, the system is then corresponding to a well-defined multipartite qudit system with local dimension $d = M$.
%In this case, the GM mode entanglement of a state is the conventional GM entanglement among distinguishable particles in qudit system.
%However, if there are multiple indistinguishable photons in local LONs, it will be intrinsic to consider GM mode entanglement among the local modes instead of the indistinguishable photons.
To reveal the physical significance of GME in multi-rail encoded LONs, one needs to evaluate the complementary properties of multiphoton states in each local LON in complementary Heisenberg-Weyl measurements \cite{WuMurao2020-CmplPropLONs}.
Like for the evaluation of bipartite multi-rail entanglement in LONs\cite{WuHofmann2017-BiEntMltMd, WuMurao2020-CmplPropLONs, KiyoharaEtAlTakeuchi2020-VerfEntBiptLONs}, one will need to derive the theoretical boundary on a GME witness or quantity for biproducible states through a convex roof extension over the irreducible subspaces of local Heisenberg-Weyl operators \footnote{The Heisenberg-Weyl operators are a generalization of Pauli operators in high-dimensional Hilbert spaces}.
%Note that, in the \whatis{rest} of this paper, we will refer to ``genuine multipartite entanglement (GME)'' as the genuine multipartite mode entanglement in the 2nd quantization formalism.

\subsection{GM-entangled states with Heisenberg-Weyl symmetry}
In multipartite qudit systems, many GM-entangled states are symmetric under simultaneous Heisenberg-Weyl (HW) transformations \cite{DurtAtElZyczkowski2010-MUBs}, e.g. particular graph states \cite{BriegelRaussendorf2001-FirstGrSt}, singlet states \cite{Cabello2002-NPtclNLvllSngltSt}, and Dicke states \cite{Dicke1954-DickeState}. The GME of these states can be detected in local complementary measurements in mutually unbiased bases \cite{SpenglerHuberEtAlHiesmayr2012-EntWitViaMUB} associated with Heisenberg-Weyl operators \cite{DurtAtElZyczkowski2010-MUBs}.
The multiphoton states that exhibit Heisenberg-Weyl symmetry are therefore of particular interest in our analysis.

In an $M$-mode LON system, a Heisenberg-Weyl operator $\widehat{X}^{i}\widehat{Z}^{j}$ is a phase shifting $\widehat{Z}^{j}$ followed by a mode shifting $\widehat{X}^{i}$, where the mode shifting $\widehat{X}$ shifts each photon to the next neighboring mode cyclicly,
\begin{equation}
\label{eq::def_mode_shift}
  \widehat{X} \widehat{a}_{m}^{\dagger} \widehat{X}^{\dagger}
  =
  \widehat{b}_{m\oplus 1}^{\dagger}
   \;\;\text{ with }\;\;
   m\oplus 1 := (m+1)_{\pmod M},
\end{equation}
and the phase shifting $\widehat{Z}$ adds a phase $\omega^{m}$ to each photon in the $m$-th mode,
\begin{equation}
\label{eq::def_phase_shift}
  \widehat{Z}\widehat{a}_{m}^{\dagger}\widehat{Z}^{\dagger} = w^{m}\widehat{b}_{m}^{\dagger}
  \;\;\text{  with  }\;\;
  \omega = e^{\imI 2\pi/M}.
\end{equation}
Here $\widehat{a}_{m}^{\dagger}$ and $\widehat{b}_{m}^{\dagger}$ are the creation operators of the $m$-th input and output modes, respectively.
The HW operators of particular interest are
\begin{equation}
  \widehat{\Lambda}_{j} := \widehat{X}\widehat{Z}^{j}
  \;\;\text{ with }\;\;
  j = 0, ..., M-1,
\end{equation}
since they can be exploited to construct mutually unbiased bases for complementary measurements.

A $P$-partite state is symmetric under simultaneous HW operation, if it is an eigenstate of the operator $\widehat{\Lambda}_{j_{1}}\otimes\cdots\otimes\widehat{\Lambda}_{j_{P}}$, where $(j_{1},...,j_{P})$ are the indices of local phase shifting. The simultaneous HW symmetry of a state can be described by two indices $(k,\kappa)$ as follows,
\begin{align}
\label{eq::HW_symmetric_state}
  \widehat{X}\otimes\cdots\otimes\widehat{X} \ket{\psi_{k,\kappa}}
  & = \omega^{k}\ket{\psi_{k,\kappa}},
  \nonumber\\
  \widehat{Z}^{j_{1}}\otimes\cdots\otimes\widehat{Z}^{j_{P}} \ket{\psi_{k,\kappa}}
  & = \omega^{\kappa}\ket{\psi_{k,\kappa}}.
\end{align}%
The state $\ket{\psi_{k,\kappa}}$ is therefore symmetric under the simultaneous HW operator,
\begin{align}
\label{eq::HW_symmetry_eigenequation}
%  \widehat{\Lambda}_{j_{1},...,j_{P}}
  \widehat{\Lambda}_{j_{1}l}\otimes\cdots \otimes\widehat{\Lambda}_{j_{P}l}
  \ket{\psi_{k,\kappa}}
  & =
  \omega^{k+\kappa l}\ket{\psi_{k,\kappa}}.
%  \text{ with }
%  \\
%  \widehat{\Lambda}_{j_{1},...,j_{P}} & :=
%  \widehat{\Lambda}_{j_{1}}\otimes\cdots \otimes\widehat{\Lambda}_{j_{P}}.
\end{align}
We call this type of symmetry the $(k,\kappa)$-symmetry.
A $(k,\kappa)$-symmetric state must be a superposition of the computational-basis states $\ket{\boldvec{n}_{1},...,\boldvec{n}_{P}}$ that satisfy $\oplus_{i}j_{i}\mu(\boldvec{n}_{i})=\kappa$,
\begin{equation}
  \ket{\psi_{k,\kappa}}
  =
  \sum_{\boldvec{n}_{i}: \oplus_{i}j_{i}\mu(\boldvec{n}_{i})=\kappa}
  c_{\boldvec{n}_{1},...,\boldvec{n}_{P}}
  \ket{\boldvec{n}_{1},...,\boldvec{n}_{P}}.
\end{equation}
Here, the quantity $\oplus_{i}j_{i}\mu(\boldvec{n}_{i})$ is the eigenphase of the $\widehat{Z}^{j_{1}}\otimes\cdots\otimes\widehat{Z}^{j_{P}}$ operator, which we call the \emph{$(j_{1},...,j_{P})$-weighted $Z$-clock label} of $\ket{\boldvec{n}_{1}, ..., \boldvec{n}_{P}}$
\begin{align}
\label{eq::def_jweighted_z-clock-lbl}
  & \widehat{Z}^{j_{1}}\otimes\cdots\otimes\widehat{Z}^{j_{P}}
  \ket{\boldvec{n}_{1}, ..., \boldvec{n}_{P}}
  \nonumber \\
  =&
  \omega^{\bigoplus_{i}j_{i} \mu(\boldvec{n}_{i})}
  \ket{\boldvec{n}_{1}, ..., \boldvec{n}_{P}}\;
  \text{ with }
  \mu(\boldvec{n}) := \sum_{m} n_{m}m.
\end{align}

Since the representation of $\widehat{\Lambda}_{j_{1}}\otimes\cdots \otimes\widehat{\Lambda}_{j_{P}}$ is reducible, the representation of a $(k,\kappa)$-symmetric state can also be reduced to a superposition of $(k,\kappa)$-symmetric states constructed within $(\widehat{\Lambda}_{j_{1}}\otimes\cdots \otimes\widehat{\Lambda}_{j_{P}})$-irreducible subclasses $\mathbb{X}$.
A $(\widehat{\Lambda}_{j_{1}}\otimes\cdots \otimes\widehat{\Lambda}_{j_{P}})$-irreducible subclass $\mathbb{X}$ is equivalent to an $\widehat{X}^{\otimes P}$-irreducible subclass, which is generated by the simultaneous mode shifting $\widehat{X}^{\otimes P}$ acting on a $P$-partite $M$-rail Fock state $\ket{\boldvec{n}_{1},...,\boldvec{n}_{P}}$,
\begin{equation}
  \mathbb{X}_{\boldvec{n}_{1},...,\boldvec{n}_{P}}
  :=
  %\{(\underset{P \text{ times}}{\underbrace{\widehat{X}\otimes\cdots\otimes\widehat{X}}})^{m}
  \left\{(\widehat{X}^{\otimes P})^{m}\ket{\boldvec{n}_{1},...,\boldvec{n}_{P}}\right\}_{m=0,...,M-1}.
\end{equation}
Within these $(\widehat{\Lambda}_{j_{1}}\otimes\cdots \otimes\widehat{\Lambda}_{j_{P}})$-irreducible subclasses, one can then construct the eigenstates of $\widehat{X}^{\otimes P}$ as a uniform superposition of the states in $\mathbb{X}$,
\begin{widetext}
\begin{equation}
\label{eq::GM_ent_component}
  \Ket{\mathbb{E}_{k}(\mathbb{X}_{\boldvec{n}_{1},...,\boldvec{n}_{P}})}
  :=
%  \frac{1}{\sqrt{d_{\mathbb{X}}}}
%  \sum_{m = 0}^{d_{\mathbb{X}}-1}
%  \omega^{-k \frac{M \, m}{d_{\mathbb{X}}}}
%  \left(\widehat{X}^{(1)}\otimes\cdots\otimes\widehat{X}^{(P)}\right)^{\frac{M\,m}{d_{\mathbb{X}}}}
%  \ket{\boldvec{n}_{1},...,\boldvec{n}_{P}}.
  \frac{\sqrt{|\mathbb{X}|}}{M}
  \sum_{m = 0}^{M-1}
  \omega^{-k m} \left(\widehat{X}\otimes\cdots\otimes\widehat{X}\right)^{m}
  \ket{\boldvec{n}_{1},...,\boldvec{n}_{P}}
  \;\;\text{ with }\;\;
%  k\in\{\frac{M}{|\mathbb{X}|} m\}_{m=0, ..., |\mathbb{X}|-1}
  k = \frac{M}{|\mathbb{X}|}, 2\frac{M}{|\mathbb{X}|},..., (|\mathbb{X}|-1)\frac{M}{|\mathbb{X}|}.
\end{equation}
\end{widetext}
The simultaneous HW operator  $\widehat{\Lambda}_{j_{1}}\otimes\cdots\otimes\widehat{\Lambda}_{j_{P}}$ will then induce a phase shift and transform an $\widehat{X}^{\otimes P}$ eigenstate as follows,
\begin{align}
  & \widehat{\Lambda}_{j_{1}}\otimes\cdots\otimes\widehat{\Lambda}_{j_{P}}
  \Ket{\mathbb{E}_{k}(\mathbb{X}_{\boldvec{n}_{1},...,\boldvec{n}_{P}})}
  \nonumber \\
  = &
  \omega^{k+\bigoplus_{i}j_{i}\mu(\boldvec{n}_{i})}
  \Ket{\mathbb{E}_{k-\bigoplus_{i}j_{i}N_{i}}(\mathbb{X}_{\boldvec{n}_{1},...,\boldvec{n}_{P}})}.
\end{align}
One can then choose the HW indices $\boldvec{j}=(j_{1},...,j_{P})$ in the following way to turn the state $\ket{\mathbb{E}_{k}(\mathbb{X})}$ into an eigenstate of the simultaneous HW operator,
\begin{equation}
\label{eq::HW_indices_cond}
  \boldvec{j}=(j_{1},...,j_{P}):
  \;\;
  \bigoplus_{i}j_{i}N_{i} = 0.
%  {\color{red}
%  \;\;\text{ and }\;\;
%  j_{i}N_{i}\neq 0 \pmod M.}
\end{equation}
%the state $\ket{\mathbb{E}_{k}(\mathbb{X})}$ is also an eigenstate of the simultaneous HW operator,
%\begin{align}
%  & \widehat{\Lambda}_{j_{1}}\otimes\cdots\otimes\widehat{\Lambda}_{j_{P}}
%  \Ket{\mathbb{E}_{k}(\mathbb{X}_{\boldvec{n}_{1},...,\boldvec{n}_{P}})}
%  \nonumber \\
%  = &
%  \omega^{k+\sum_{i}j_{i}\mu(\boldvec{n}_{i})}
%  \Ket{\mathbb{E}_{k}(\mathbb{X}_{\boldvec{n}_{1},...,\boldvec{n}_{P}})}
%\end{align}
The eigenstate $\ket{\mathbb{E}_{k}(\mathbb{X})}$ in Eq. \eqref{eq::GM_ent_component} is therefore $(k,\kappa)$-symmetric with $\kappa = \bigoplus_{i}j_{i}\mu(\boldvec{n}_{i})$.

In this case, every element of an $\mathbb{X}$ class has the same $\boldvec{j}$-weighted $Z$-clock label, which we call the $\boldvec{j}$-weighted $Z$-clock label of a class $\mathbb{X}$,
\begin{equation}
  \mu_{\boldvec{j}}(\mathbb{X})
  :=
  \bigoplus_{i} j_{i}\mu(\boldvec{n}_{i})
  \text{ with }
  \ket{\boldvec{n}_{1},...,\boldvec{n}_{P}}\in\mathbb{X}.
\end{equation}
As a result, a state $\ket{\psi_{k,\kappa}}$ that exhibits the $(k,\kappa)$-symmetry described in Eq. \eqref{eq::HW_symmetric_state} can be expressed as a superposition of the $\widehat{\Lambda}_{j_{1}}\otimes\cdots\otimes\widehat{\Lambda}_{j_{P}}$ eigenstates $\Ket{\mathbb{E}_{k}(\mathbb{X})}$ over all $\widehat{X}^{\otimes P}$-irreducible subclasses, of which the $(j_{1},...,j_{P})$-weighted $Z$-clock label is $\kappa$
\begin{equation}
\label{eq::HW-symmetric_st}
  \ket{\psi_{k,\kappa}}
  =
  \sum_{
    \mu_{\boldvec{j}}(\mathbb{X}) = \kappa
  }c_{\mathbb{X}}\Ket{\mathbb{E}_{k}(\mathbb{X})}.
\end{equation}
Mathematically, this state is highly GM-entangled.
A general $(N_{1},...,N_{P})$-photon pure state can be written as a superposition of HW symmetric states
\begin{equation}
\label{eq::repres_in_kKappa}
  \ket{\Phi_{N_{1},...,N_{P}}} =
  \sum_{k,\kappa}c_{k,\kappa}\ket{\psi_{k,\kappa}}.
\end{equation}
The more unbalanced the superposition in Eq. \eqref{eq::repres_in_kKappa} is, the greater the GME is.
The extremum case is $c_{k,\kappa}=1$, which leads to a unique HW symmetry.
One can therefore pick the HW-symmetric component with the highest probability amplitude $|c_{k,\kappa}|$ as the target GM-entangled state and construct a corresponding GME detection measurement to verify the GME of $\ket{\Phi_{N_{1},...,N_{P}}}$.

\subsection{Detection of GME}

\begin{figure*}[t]
  \centering
  \hfill
  \subfloat[]{\includegraphics[width=0.45\textwidth]{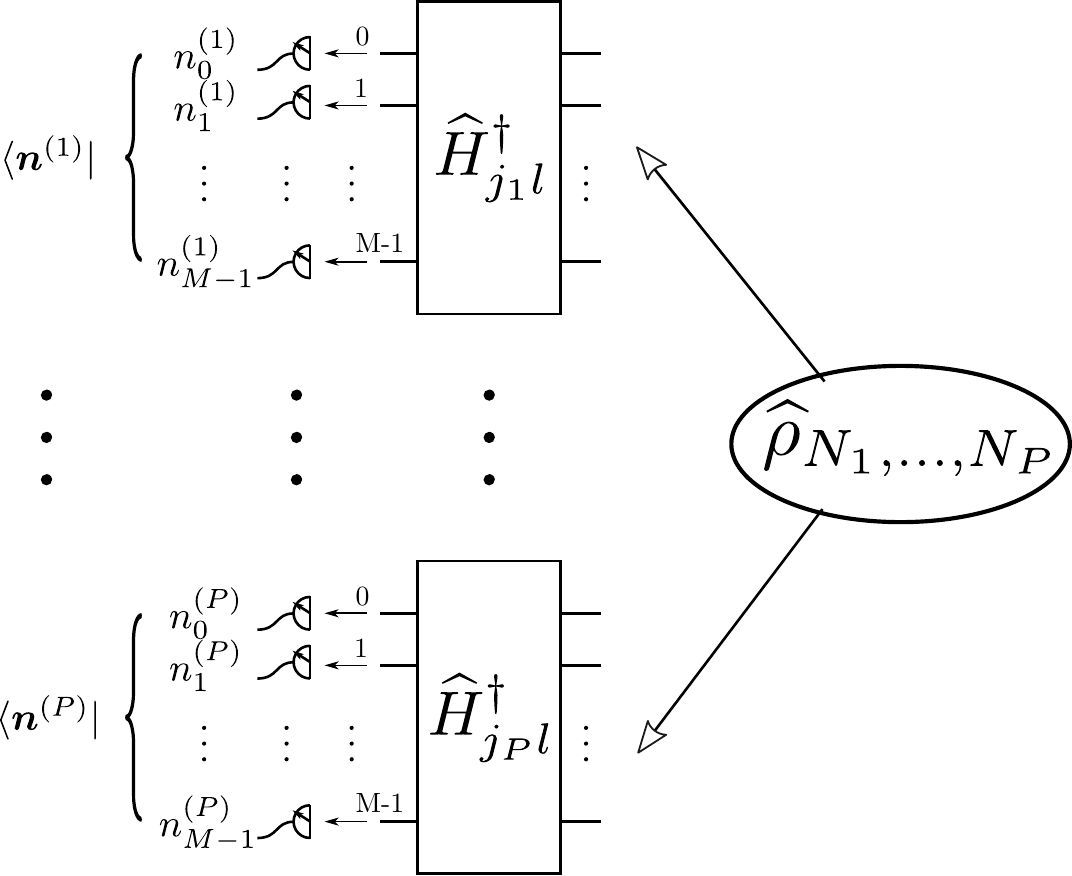}}
  \hfill
  \subfloat[]{\includegraphics[width=0.4\textwidth]{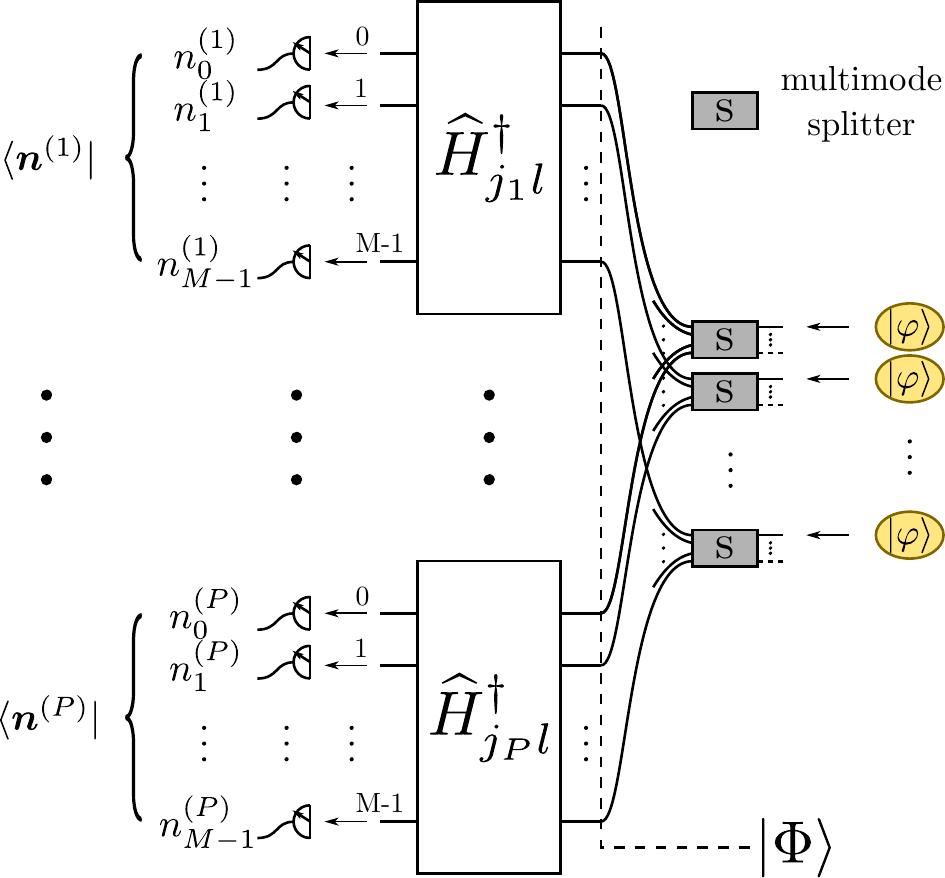}}
  \hfill
  \caption{%
    (a) Measurement in the $\widehat{\Lambda}_{j_{1}l}\otimes\cdots\otimes\widehat{\Lambda}_{j_{P}l}$ eigenbasis.
    (b) Generation and evaluation of genuine $P$-partite entanglement employing $P$-mode splitters.
  }\label{fig::GEM_gen_dect}
\end{figure*}

The entanglement of an HW-symmetric state is the target GME that we want to verify with the theory developed in this section.
Each component $\ket{\mathbb{E}_{k}(\mathbb{X})}$ of an HW-symmetric state in Eq. \eqref{eq::HW-symmetric_st} is a highly GM-entangled state.
As an intrinsic property of GME, the HW symmetry is a strong indication of GME, if one can reveal its characteristic complementary correlations in the eigenbases of corresponding local HW operators that are mutually unbiased to each other.
%To this end, one will need local complementary measurements
% to evaluate the correlations of a testing in complementary bases (i.e. mutually unbiased bases).
In such local complementary measurements, one can determine the upper bound on the complementary correlations for biproducible states through the convex roof extension of the upper bounds for different $\widehat{X}^{\otimes P}$-irreducible subspaces \cite{WuMurao2020-CmplPropLONs}.

The complementary correlations will be evaluated through a GME verifier that stabilizes a target HW-symmetric state \cite{Wu2020-AdptQSFEBiEnt}.
It can be constructed as a projection onto the supporting outputs of corresponding HW measurement settings.
The projection onto the subspace that exhibits symmetry with an eigenphase of $m$ under the HW transformation $\widehat{\Lambda}_{j_{1}l}\otimes\cdots\otimes\widehat{\Lambda}_{j_{P}l}$ can be constructed as
\begin{align}
\label{eq::S_Lambda_def_0}
  \widehat{S}_{\Lambda|(l,m)}
  & :=
  \frac{1}{M}\sum_{m'=0}^{M-1} \omega^{-m m'}
  \widehat{\Lambda}_{j_{1}l}^{m'}
  \otimes \cdots \otimes
  \widehat{\Lambda}_{j_{P}l}^{m'}.
\end{align}
This projection is a sum of the projectors $\projector{\mathbb{E}_{k}(\mathbb{X})}$ in the corresponding $\widehat{X}^{\otimes P}$-irreducible classes $\mathbb{X}$, of which the $\boldvec{j}$-weighted $Z$-clock label satisfies $k\oplus\mu_{\boldvec{j}}(\mathbb{X})l = m$,
\begin{align}
\label{eq::S_Lambda_def_1}
  \widehat{S}_{\Lambda|(l,m)}
  =
  \sum_{\mathbb{X}: k\oplus\mu_{\boldvec{j}}(\mathbb{X})l = m}
  \Projector{\mathbb{E}_{k}(\mathbb{X})}.
\end{align}
As a result of Eqs. \eqref{eq::HW-symmetric_st} and \eqref{eq::S_Lambda_def_1}, the projector $\widehat{S}_{\Lambda|(l,m)}$ with $m=k+\kappa l$ stabilizes a $(k,\kappa)$-symmetric state $\ket{\psi_{k,\kappa}}$,
\begin{equation}
  \widehat{S}_{\Lambda|(l, m)} \ket{\psi_{k,\kappa}}
  =
  \delta_{k+\kappa l}^{m} \ket{\psi_{k,\kappa}}.
\end{equation}

According to Theorem 3.1 in \cite{WuMurao2020-CmplPropLONs}, the projector $\widehat{S}_{\Lambda|(k, m)}$ can be evaluated in the the $(\widehat{\Lambda}_{j_{1}l}\otimes \cdots \otimes \widehat{\Lambda}_{j_{P}l})$-eigenbasis measurement, which is implemented with the local generalized Hadamard transformation $\widehat{H}_{j_{1}l}\otimes\cdots\otimes\widehat{H}_{j_{P}l}$ (see Fig. \ref{fig::GEM_gen_dect} (a)),
\begin{widetext}
\begin{align}
\label{eq::S_Lambda_msmnt}
  \widehat{S}_{\Lambda|(l,m)}
  %\nonumber \\
  =
  \widehat{H}_{j_{1}l}\otimes\cdots\otimes\widehat{H}_{j_{P}l}
  \left(\sum_{\oplus_{i}\mu(\boldvec{n}_{i}) = m}
  \projector{\boldvec{n}_{1},...,\boldvec{n}_{P}}\right)
  \widehat{H}_{j_{1}l}^{\dagger}\otimes\cdots\otimes\widehat{H}_{j_{P}l}^{\dagger},
\end{align}
\end{widetext}
where $\widehat{H}_{j}$ is a generalized Hadamard transformation
\begin{equation}
  \widehat{H}_{j}^{\dagger}\widehat{a}^{\dagger}_{m}\widehat{H}_{j}
  =
  \frac{1}{\sqrt{M}}\omega^{\frac{1}{2}(M-m)mj}
  \sum_{m'}\omega^{m'm}\widehat{b}_{m}^{\dagger}.
\end{equation}
For $j=0$, $\widehat{H}_{0}$ is the discrete Fourier transformation.
The projector $\widehat{S}_{\Lambda|(l,m)}$ therefore projects input states onto the $\widehat{\Lambda}_{j_{1}l}\otimes\cdots\otimes\widehat{\Lambda}_{j_{P}l}$ eigensubspace that has a correlation of $\bigoplus_{i}\mu(\boldvec{n}_{i}) = m$.

On the other hand, in the computational basis, one can also construct a stabilizer of a $(k,\kappa)$-symmetric state
\begin{equation}
\label{eq::Sz_k_kappa}
  \widehat{S}_{Z| m }
  =
  \sum_{
    \oplus_{i}j_{i}\mu(\boldvec{n}_{i}) = m
  }
  \projector{\boldvec{n}_{1},...,\boldvec{n}_{P}},
\end{equation}
such that
\begin{equation}
  \widehat{S}_{Z| m } \ket{\psi_{k,\kappa}}
  =
  \delta_{\kappa}^{m} \ket{\psi_{k,\kappa}}.
\end{equation}

Mixing the two types of state stabilizers in Eqs. \eqref{eq::S_Lambda_msmnt} and \eqref{eq::Sz_k_kappa}, one can construct a GME verifier to verify the GME that exhibits a HW symmetry associated with the indices $(k,\kappa)$,
\begin{equation}
\label{eq::GME_verifier}
  \widehat{V}_{k,\kappa}
  :=
  \frac{1}{1+|\mathbb{L}|}
  \left(
    \widehat{S}_{Z|\kappa}
    +
    \sum_{l\in\mathbb{L}}
    \widehat{S}_{\Lambda|(l,k+\kappa l)}
  \right)
\end{equation}
where the configuration set $\mathbb{L}$ of the $\bigotimes_{i}\widehat{\Lambda}_{j_{i}l}$ measurements can be chosen as a subset of $\{0,1,...,M-1\}$.
It is obvious that the GME verifier $\widehat{V}_{k,\kappa}$ stabilizes a $(k,\kappa)$-symmetric state $\ket{\psi_{k,\kappa}}$,
\begin{equation}
  \widehat{V}_{k,\kappa}\ket{\psi_{k',\kappa'}}
  =
  \frac{1}{1+|\mathbb{L}|}
  \left(\delta_{\kappa'}^{\kappa} + \sum_{l\in\mathbb{L}}\delta_{k'+\kappa'l}^{k+\kappa l}\right)
  \ket{\psi_{k',\kappa'}}.
\end{equation}

The configuration $\mathbb{L}$ must be chosen in a way such that the $\bigotimes_{i}\widehat{\Lambda}_{j_{i}l}$ measurements are complementary to each other, i.e. they have mutually unbiased measurement bases.
According to Theorem 2.2 in \cite{WuMurao2020-CmplPropLONs}, such a complementary configuration set $\mathbb{L}$ must fulfill
\begin{equation}
\label{eq::msmnt_set_cond}
%  \gcd\left(j_{i}(l-l')N_{i} d_{\mathbb{X}_{i}} /M, d_{\mathbb{X}_{i}}\right) = 1
  \gcd\left(j_{i}(l-l')N_{i} |\mathbb{X}_{\boldvec{n}_{i}}| /M, |\mathbb{X}_{\boldvec{n}_{i}}|\right) = 1
\end{equation}
for all local systems $i\in\{1,...,P\}$, $l,l'\in\mathbb{L}$, and
\begin{equation}
  \braket{\boldvec{n}_{1},...,\boldvec{n}_{P}|\widehat{\rho}|\boldvec{n}_{1},...,\boldvec{n}_{P}} \neq 0.
\end{equation}
The choice of $\mathbb{L}$ is adaptive to the measurement statistics in the computational basis.
The construction of measurement settings according to Eq. \eqref{eq::msmnt_set_cond} guarantees that the local eigenbases of $\widehat{\Lambda}_{j_{1}l}\otimes\cdots\otimes\widehat{\Lambda}_{j_{P}l}$ measurements with $l\in\mathbb{L}$ are mutually unbiased within all the local $\widehat{X}$-irreducible subspaces
$\bigotimes_{i}\spn(\mathbb{X}_{i})$ that supports the testing state $\widehat{\rho}$.
As a result of the complementarity of the local measurement, one can then detect the GME according to the following theorem.
%\begin{widetext}
\begin{theorem}[GME detection]
\label{theorem::GME_detection}
  For a target HW-symmetric state $\ket{\psi_{k,\kappa}}$ defined in Eq. \eqref{eq::HW_symmetric_state} with fixed local photon numbers $\{N_{1},...,N_{P}\}$, one can construct a set of measurements implemented by the local generalized Hadamard transform $\widehat{H}_{j_{1}\,l}\otimes\cdots\otimes\widehat{H}_{j_{P}\,l}$, where the indices $\{j_{1},...,j_{P}\}$ satisfy Eq. \eqref{eq::HW_indices_cond}, and $l\in\mathbb{L}$ satisfies Eq. \eqref{eq::msmnt_set_cond}.
  In this set of measurements, one can construct a GME verifier $\widehat{V}_{k,\kappa}$ through Eq. \eqref{eq::GME_verifier}.
  The upper bound on the expectation value $\braket{\widehat{V}_{k,\kappa}}$ for bi-producible states is
%  \begin{equation}
%  \label{eq::thm_sep_bound_general}
%    \tr\left(\widehat{V}_{k,\kappa}\widehat{\rho}\right)
%    \le
%    \frac{1}{|\mathbb{L}|+1}
%    \left(
%      1+
%      |\mathbb{L}|\sum_{\mathbb{X}_{1},...,\mathbb{X}_{P}}
%      \frac{p_{\mathbb{X}_{1},...,\mathbb{X}_{P}}(\rho)}{d[\mathbb{X}_{1},...,\mathbb{X}_{P}]}
%    \right),
%    \text{ for all bi-producible }
%    \widehat{\rho}.
%  \end{equation}
  \begin{equation}
  \label{eq::thm_sep_bound_general}
    %\tr\left(\widehat{V}_{k,\kappa}\widehat{\rho}\right)
    \braket{\widehat{V}_{k,\kappa}}
    \le
    \frac{1+\braket{\widehat{D}} |\mathbb{L}|}{1+|\mathbb{L}|},
    \text{ for all bi-producible }
    \widehat{\rho},
  \end{equation}
  where $\widehat{D}$ is an operator evaluated in the computational basis
  %the average of {a quantity evaluated in the computational basis},
%  that evaluates the minimum correlations averaged over the
%  average bipartite correlation \whatis{in the ... basis under maximum randomization}
  \begin{equation}
    \widehat{D} =
    \sum_{\boldvec{n}_{i}: |\boldvec{n}_{i}|=N_{i}}
    \frac{ 1 }{
      \min_{i}|\mathbb{X}_{\boldvec{n}_{i}}|
    }
    \projector{\boldvec{n}_{1},...,\boldvec{n}_{P}},
  \end{equation}
  and $\min_{i}|\mathbb{X}_{\boldvec{n}_{i}}|$ is
  the minimum cardinality of local $\widehat{X}$-irreducible classes $\mathbb{X}_{\boldvec{n}_{i}}$ for the Fock vectors $(\boldvec{n}_{1}, ..., \boldvec{n}_{P})$.
\begin{proof}
  See Appendix \ref{sec::proof}.
\end{proof}
\end{theorem}%
%\end{widetext}
If the expectation value $\braket{\widehat{V}_{k,\kappa}}$ exceeds this bound, then one can conclude the GME of the testing state.
It is obvious that a $(k,\kappa)$-symmetric state $\ket{\psi_{k,\kappa}}$ has the maximum expectation value,
\begin{equation}
\label{eq::V_exp_kKappa}
  \braket{\psi_{k,\kappa}|\widehat{V}_{k,\kappa}|\psi_{k,\kappa}}
  =
  1,
\end{equation}
which exceed the upper bound for bi-producible states determined in Eq. \eqref{eq::thm_sep_bound_general}.

Note that the choice of $\mathbb{L}$ depends on local photon numbers $N_{i}$, mode number $M$, and HW indices $\boldvec{j}$ according to Eq. \eqref{eq::msmnt_set_cond}. If one chooses $\mathbb{L} = \{0\}$, the condition in Eq. \eqref{eq::msmnt_set_cond} is always fulfilled. The single-index set $\mathbb{L} = \{0\}$ is always a good choice for GME detection. In this construction, one can detect GME with just two measurement settings, one is in the computational basis, and the other is in the $\widehat{X}^{\otimes P}$ eigenbasis implemented by the inverse discrete Fourier transformation $\widehat{H}_{0}^{\dagger \otimes P}$.

For the special case when $M$ is a prime number, the cardinality of all local $\widehat{X}$-irreducible classes $\mathbb{X}_{\boldvec{n}_{i}}$ is $M$. Meanwhile, Eq. \eqref{eq::msmnt_set_cond} is always satisfied, if the local photon number $N_{i}$ is not a multiple of $M$. One can therefore choose an arbitrary subset of $\{0,...,M-1\}$ as the settings $\mathbb{L}$ for complementary measurements.
In this case, Theorem \ref{theorem::GME_detection} can be simplified as the following corollary.%
\begin{corollary}
\label{coro::GME_detection_prime}
  For a multipartite LONs, in which each local mode number $M$ is prime, one can construct a set of measurements implemented by the local generalized Hadamard transform $\widehat{H}_{j_{1}\,l}\otimes\cdots\otimes\widehat{H}_{j_{P}\,l}$, where the indices $\{j_{1},...,j_{P}\}$ satisfy Eq. \eqref{eq::HW_indices_cond}, and $l\in\mathbb{L}$ with $\mathbb{L}\subseteq\{0, ..., M-1\}$.
  The GME verifier $\widehat{V}_{k,\kappa}$ given in Eq. \eqref{eq::GME_verifier} then has the upper bound
  \begin{equation}
  \label{eq::coro_sep_bound_prime}
    \tr\left(\widehat{V}_{k,\kappa}\widehat{\rho}\right)
    \le
    \frac{M+|\mathbb{L}|}{M(|\mathbb{L}|+1)},
    \text{ for all bi-producible }
    \widehat{\rho}.
  \end{equation}
\begin{proof}
  For a prime number $M$, any $l\in\{0,...,M-1\}$ satisfies Eq. \eqref{eq::msmnt_set_cond}, and one can therefore choose any $l$ to construct the set of complementary configuration $\mathbb{L}$.
  In this case, the expectation value of $\widehat{D}$ in Theorem \ref{theorem::GME_detection} is always equal to $1/M$, which leads to the upper bound given in \eqref{eq::coro_sep_bound_prime}.
\end{proof}
\end{corollary}

If one chooses the complete complementary set $\mathbb{L} = \{0,...,M-1\}$, the bound is simply $2/(M+1)$.
In such measurement settings, the expectation value of $\widehat{V}_{k,\kappa}$ for a general $(N_{1},...,N_{P})$-photon pure state $\ket{\Phi_{N_{1},...,N_{P}}}$ in Eq. \eqref{eq::repres_in_kKappa} is given by
\begin{equation}
  \braket{\Phi_{N_{1},...,N_{P}}|\widehat{V}_{k,\kappa}|\Phi_{N_{1},...,N_{P}}}
  =
  \frac{1 + M\,|c_{k,\kappa}|^{2}}{M+1}.
\end{equation}
As a result, for a prime-number $M$, one can always detect the GME for a state $\ket{\Phi_{N_{1},...,N_{P}}}$, which has a $(k,\kappa)$ probability amplitude larger than $1/\sqrt{M}$,
\begin{equation}
\label{eq::prime-number_criterion}
  |c_{k,\kappa}|^{2} > \frac{1}{M}
  \Rightarrow
  \text{ GME }.
\end{equation}
Let us consider a state $\ket{\Phi_{N_{1},...,N_{P}}^{(k)}}$ with a particular $\omega^{k}$ eigenphase symmetry under the $\widehat{X}^{\otimes P}$ transformation
\begin{equation}
  \ket{\Phi_{N_{1},...,N_{P}}^{(k)}}
  =
  \sum_{\kappa = 0}^{M-1} c_{k,\kappa} \ket{\psi_{k,\kappa}}.
\end{equation}
It always has a maximum probability amplitude $|c_{\tilde{k},\tilde{\kappa}}|^{2}>1/M$, unless $|c_{k,\kappa}|^{2} = 1/M$ for all $\kappa$.
This means that we have a very high probability to detect GME of $\ket{\Phi_{N_{1},...,N_{P}}^{(k)}}$ with our measurement settings.
For GME generation, it is therefore desirable to create a state which is symmetric under the simultaneous mode shifting $\widehat{X}^{\otimes P}$.

\section{Generation of multipartite $M$-rail entanglement}
\label{sec::generation}
In this section, we consider the generation of GME in multipartite $M$-rail systems through state preparation for $\widehat{X}^{\otimes P}$-symmetry.
A generation scheme for multipartite $M$-rail states that exhibit $\widehat{X}^{\otimes P}$-symmetry can be extended from the scheme for bipartite $M$-rail entanglement in \cite{WuHofmann2017-BiEntMltMd, KiyoharaEtAlTakeuchi2020-VerfEntBiptLONs}.

As shown in Fig. \ref{fig::GEM_gen_dect} (b), $M$ copies of input state $\ket{\varphi}^{\otimes M}$ are sent into $M$ pieces of multimode splitters with $P$ output modes.
A $P$-mode splitter divides the $m$th-input into a uniform superposition of $M$ modes distributed in each local system indexed by the label $m$,
\begin{equation}
  \widehat{S}\,\widehat{a}_{m}^{\dagger}\,\widehat{S}^{\dagger}
  =
  \frac{1}{\sqrt{P}} (\widehat{a}_{1,m}^{\dagger} + \cdots + \widehat{a}_{P,m}^{\dagger}),
\end{equation}
where $\widehat{a}_{i,m}^{\dagger}$ is the creation operator of the $m$-th mode in the $i$-th local system.
In general, a single-mode input $\ket{\varphi}$ can be expressed in the 2nd quantization formalism as
\begin{equation}
  \ket{\varphi} = \sum_{\nu}c_{\varphi}(\nu)\ket{\nu},
\end{equation}
where $\ket{\nu}$ are Fock states.
If we have $M$ copies of this state, the input state is a superposition of different Fock-state vectors $\ket{\boldvec{\nu}} := \ket{\nu_{0},...,\nu_{M-1}}$,
\begin{equation}
  \ket{\varphi}^{\otimes M} = \sum_{\boldvec{\nu}}c_{\varphi}(\boldvec{\nu})\ket{\boldvec{\nu}},
\end{equation}
where $c_{\varphi}(\boldvec{\nu})$ is the product of the probability amplitude $c_{\varphi}(\nu_{m})$ at each mode
\begin{equation}
  c_{\varphi}(\boldvec{\nu}) := \prod_{m=0}^{M-1}c_{\varphi}(\nu_{m}).
\end{equation}
The $(N_{1},...,N_{P})$-postselected state $\Ket{\Phi_{N_{1},...,N_{P}}^{}(\varphi)}$ is a superposition of Fock-vector states $\ket{\boldvec{n}_{1},...,\boldvec{n}_{P}}$ that have local photon numbers $|\boldvec{n}_{i}|=N_{i}$,
\begin{align}
\label{eq::postsel_psi}
  & \ket{\Phi^{}_{N_{1},...,N_{P}}(\varphi) } = \frac{1}{\sqrt{p^{}_{N_{1},...,N_{P}}}}\times
  \nonumber \\
  %= &
%  \frac{1}{\sqrt{p^{}_{N_{1},...,N_{P}}}}
  &
  \sum_{\boldvec{n}_{i}:|\boldvec{n}_{i}|=N_{i}}
  \frac{
    c_{\varphi}( \boldvec{n}_{\text{tot}})
  }{
    \sqrt{P^{N_{\text{tot}}}}
  }
  \sqrt{
    \frac{\boldvec{n}_{\text{tot}}!}{\boldvec{n}_{1}!\cdots\boldvec{n}_{P}!}
  }
  \ket{\boldvec{n}_{1},...,\boldvec{n}_{P}}.
\end{align}
Here, $N_{\text{tot}}:=\sum_{i}N_{i}$ is the total photon number, $\boldvec{n}_{\text{tot}} :=\sum_{i}\boldvec{n}_{i}$ is the Fock vector component of the input before the multimode splitters, and $p^{}_{N_{1},...,N_{P}}$ is the probability of the postselction on the local photon numbers
$(N_{1},...,N_{P})$,
\begin{align}
  p^{}_{N_{1},...,N_{P}}
  =
  \sum_{\boldvec{n}_{i}:|\boldvec{n}_{i}|=N_{i}}
  \frac{|c_{\varphi}(\boldvec{n}_{\text{tot}})|^{2}}{P^{N_{\text{tot}}}}
  \frac{\boldvec{n}_{\text{tot}}!}{\boldvec{n}_{1}!\cdots\boldvec{n}_{P}!}.
\end{align}
It is obvious that the state $\ket{\Phi_{N_{1},...,N_{P}}(\varphi)}$ is $\widehat{X}^{\otimes P}$-symmetric,
\begin{equation}
  \widehat{X}^{\otimes P}\ket{\Phi_{N_{1},...,N_{P}}(\varphi)}
  =
  \ket{\Phi_{N_{1},...,N_{P}}(\varphi)}.
\end{equation}
It is therefore a superposition of $(0,\kappa)$-symmetric states
\begin{equation}
\label{eq::0sym_st}
  \ket{\Phi_{N_{1},...,N_{P}}(\varphi)}
  = \sum_{\kappa} c_{\kappa} \ket{\psi_{0,\kappa}}.
\end{equation}

According to Eq. \eqref{eq::prime-number_criterion}, in a prime-number $M$-mode system, for such an $\widehat{X}^{\otimes P}$ symmetric state, the only case in which our method cannot detect GME of $\ket{\Phi_{N_{1},...,N_{P}}(\varphi)}$ is when $|c_{\kappa}|^{2} = 1/M$ for all $\kappa \in \{0,...,M-1\}$.
Such a uniform superposition with $|c_{0,\kappa}| = 1/\sqrt{M}$ can be generated with coherent states $\ket{\varphi} = \ket{\alpha}$ as inputs.
In this case, the state $\ket{\Phi_{N_{1},...,N_{P}}(\alpha)}$ is fully separable.
The expectation value $\braket{\widehat{V}_{0,\kappa}}$ achieves its maximum for bi-producible states.
\begin{equation}
  \braket{\Phi_{N_{1},...,N_{P}}(\alpha)|\widehat{V}_{0,\kappa}|\Phi_{N_{1},...,N_{P}}(\alpha)}
  =
  \frac{2}{M+1}
\end{equation}
for all $\kappa$.
The more unbalanced the superposition in Eq. \eqref{eq::0sym_st} is, the higher the GME of $\ket{\Phi_{N_{1},...,N_{P}}(\varphi)}$ is.
With single-photon and squeezed-state inputs, one can always generate an $\widehat{X}^{\otimes P}$-symmetric state with an unbalanced superposition in $\kappa$.

\subsection{Generation of GME with single-photon input sources}
\label{sec::GME_gen_1ph}
For single-photon inputs $\ket{\varphi} = \ket{1}$, the $(N_{1},...,N_{P})$-post-selected state of GME generation in Fig. \ref{fig::GEM_gen_dect} (b) is
\begin{equation}
  \ket{\Phi_{N_{1},...,N_{P}}}
  =
%  \sqrt{\frac{N_{1}!\cdots N_{P}!}{M!}}
  \frac{1}{\sqrt{\binom{M}{N_{1},...,N_{P}}}}
  \sum_{\substack{
    \boldvec{n}_{i}:|\boldvec{n}_{i}|=N_{i}
    \\
    \Sigma_{i}\boldvec{n}_{i} = (1,...,1)
  }}
%  \sum_{\Sigma_{i}\boldvec{n}_{i} = (1,...,1)}^{\boldvec{n}_{i}:|\boldvec{n}_{i}|=N_{i}}
  \ket{\boldvec{n}_{1},...,\boldvec{n}_{P}}.
  %\ket{\boldvec{n}_{1},...,\boldvec{n}_{P}}.
\end{equation}
To detect the GME of this state, one can choose the HW-indices $\boldvec{j} = (1,...,1)$, which fulfill Eq. \eqref{eq::HW_indices_cond}, to construct the GME detection measurements.
The state $\ket{\Phi_{N_{1},...,N_{P}}}$ exhibits the $(k,\kappa)$-symmetry with $k=0$ and $\kappa=M(M-1)/2$,
\begin{align}
  \widehat{X}^{\otimes P} \ket{\Phi_{N_{1},...,N_{P}}}
  & = \ket{\Phi_{N_{1},...,N_{P}}},
  \\
  \widehat{Z}^{\otimes P} \ket{\Phi_{N_{1},...,N_{P}}}
  & = \omega^{M(M-1)/2} \ket{\Phi_{N_{1},...,N_{P}}}.
\end{align}

For a general mode number $M$, which has $d_{M}$ as its minimum prime-number divider, one can construct the complementary measurements for GME detection according to Theorem \ref{theorem::GME_detection}.
Since $l\in\{0,...,d_{M}-1\}$ fulfills Eq. \eqref{eq::msmnt_set_cond}, one can choose any subset $\mathbb{L}\subseteq\{0,...,d_{M}-1\}$ as the set of complementary measurement configurations.
%that can include additional measurements in the $\widehat{\Lambda}_{l}$-eigenbasis with $\mathbb{L}\subseteq\{0,...,d_{M}-1\}$ as the set of complementary measurement configurations.
The GME verifier in Eq. \eqref{eq::GME_verifier} associated with these measurement settings is then given by
\begin{equation}
  \widehat{V}_{0,\kappa}
  =
  \frac{1}{1+|\mathbb{L}|}
  \left(
    \widehat{S}_{Z|(0, \kappa)}
    +
    \sum_{l\in \mathbb{L}}\widehat{S}_{\Lambda|(0,\kappa l)}
  \right),
\end{equation}
with $\kappa = M(M-1)/2$.
The expectation value of $\widehat{V}_{0,\kappa}$ for the state $\ket{\Phi_{N_{1},...,N_{P}}}$ is unity
\begin{equation}
\label{eq::1Ph_GME_detection}
  \braket{\Phi_{N_{1},...,N_{P}}|\widehat{V}_{0,\kappa}|\Phi_{N_{1},...,N_{P}}}
  =
  1 >
  \beta_{\text{biprod.}},
\end{equation}
where the upper bound $\beta_{\text{biprod.}}$ for bi-producible states can be determined by Eq. \eqref{eq::thm_sep_bound_general}.
If the mode number $M$ is a prime number, the upper bound is simply determined by Eq. \eqref{eq::coro_sep_bound_prime}.

In a more general scheme, one can also employ Fock states $\ket{\nu}$ with photon number $\nu$ larger than $1$ as the inputs in the GME generation shown in Fig. \ref{fig::GEM_gen_dect} (b).
If $\nu$ is not a divider of $M$, the corresponding measurement construction of GME detection is identical to the one constructed for single-photon inputs.
On the other hand, if $\nu$ is a divider of $M$, Eq. \eqref{eq::msmnt_set_cond} does not hold for any $l\neq0$.
In this case, the GME detection measurement can be constructed only with the measurement in the computational basis and an additional measurement in the $\widehat{X}^{\otimes P}$ eigenbasis.

From Eq. \eqref{eq::1Ph_GME_detection}, one can see that indistinguishable Fock-state inputs result in perfect HW symmetric states, which have the optimum GME signature.
In addition, another advantage of GME generation with Fock-state inputs is the robustness of GME generation against photon losses, since the postselection of the measurement outputs on local photon numbers $(N_{1},...,N_{P})$, for which the total photon number $\sum_{i}N_{i}$ is equal to the input total photon number, already excludes the photon losses in the statistics.
However, the scalability of such a generation scheme is a challenging issue in practice due to the difficulty of preparing indistinguishable single-photon sources.

\subsection{Generation of GME with displaced squeezed vacuum}

A solution for the scalability of the GME generation scheme in Fig. \ref{fig::GEM_gen_dect}(b) is to replace the single-photon inputs with CV sources, such as squeezed vacuums. In addition, we introduce displacement on the input squeezed vacuums to reveal the change in the GME signature in low-photon-number subspaces. Since adding displacement on each input mode is equivalent to a local displacement operation on each output mode, it does not change the overall GME of the whole CV state. However, as the displacement increases, the photon statistics of low-photon-number subspaces tends to behave as a coherent state. One will therefore expect a diminishing of the corresponding GME in these subspaces, which indicates a flow of GME from lower-photon-number subspaces to higher-photon-number subspaces. In this section, we reveal the GME signature in these low-photon-number subspaces using our GME verifier.

The state we consider is the $x$-displaced $r$-squeezed vacuum state $\ket{\varphi} = \ket{\sigma(r,x)}$ with
\begin{equation}
  \ket{\sigma(r,x)}
  =
  (1-\gamma^{2})^{\frac{1}{4}}e^{-\frac{2\gamma (\zeta x)^{2} }{1+\gamma}}
  \sum_{n}
%  \left(
  \sqrt{
    \frac{\gamma^{n}}{n!}
  }
%  \right)^{\frac{1}{2}}
  h_{n}(2\zeta x)
  \ket{n},
\end{equation}
where $h_{n}$ is the probabilists' Hermite polynomial, and $\gamma$ and $\zeta$ are determined by the squeezing factor $r$\footnote{The squeezing factor $r$ corresponds to a squeezed quadrature uncertainty $\Delta x = \exp(-r)\Delta x_{coh}$, where $\Delta x_{coh}$ is the quadrature uncertainty of a coherent state.}
%is converted to the unit ``dB'' through $r_{dB} = (20\log_{10}e)\,r$.
%\begin{equation}
%  \gamma = \frac{1-r}{1+r}
%  \text{ and }
%  \zeta = \sqrt{\frac{1}{1-r^{2}}}.
%\end{equation}
\begin{equation}
  \gamma = \tanh(r)
  \;\;\text{ and }\;\;
  \zeta = \sqrt{\frac{1}{1-e^{-4r}}}.
\end{equation}
According to Eq. \eqref{eq::postsel_psi}, the $(N_{1},...,N_{P})$-postselected quantum state in this generation scheme is
\begin{align}
\label{eq::ent_st_from_sq_input}
  & \ket{\Phi^{}_{N_{1},...,N_{P}}(r,x)}
  \nonumber \\
  = &
  \frac{1}{\sqrt{R_{N_{1},...,N_{P}}}}
  \sum_{\boldvec{n}_{i}:|\boldvec{n}_{i}|=N_{i}}
  \frac{
    h_{\boldvec{n}_{\text{tot}}}(2\zeta x)
  }{
    \sqrt{\boldvec{n}_{1}!\cdots\boldvec{n}_{P}!}
  }
  \ket{\boldvec{n}_{1},...,\boldvec{n}_{P}}
\end{align}
where $h_{\boldvec{n}}(x) = \prod_{m} h_{n_{m}}(x)$ is a product of the Hermite polynomials $h_{n_{m}}(x)$, and
%\begin{equation}
%  h_{\boldvec{n}}(x)
%  =
%  \prod_{m} h_{n_{m}}(x)
%\end{equation}
$R_{N_{1},...,N_{P}}$ is a normalization factor.

In Eq. \eqref{eq::ent_st_from_sq_input}, one can see that the squeezing factor $r$  affects only the factor $\zeta$, which changes only the scaling of the displacement $x$.
Changing the squeezing factor $r$ therefore does not change the behavior of the expectation value of the GME verifier $\braket{\widehat{V}_{0,\kappa}}$ against the displacement $x$.
For a large enough $r$, the scaling factor is approximately given by
\begin{equation}
  \zeta(r) \approx 1 + \frac{1}{2} e^{-4r}.
\end{equation}
For $r \gtrapprox 0.576 \; (5 dB)$, the scaling factor $\zeta-1\lessapprox 0.05$, which means increasing the squeezing factor $r$ does not introduce a significant change in the post-selected state anymore.
However, one should note that it still affect the efficiency of the post-selection on $(N_{1},...,N_{P})$ local photon numbers.

If the squeezing factor $r=0$, the input states are simply coherent states, which will create a fully separable state.
Since the photon statistics of a displaced squeezed vacuum is approximately equal to the photon statistics of coherent states for a large displacement, one expects the GME signature detected by the GME verifier $\widehat{V}_{0,\kappa}$ to asymptotically diminish as the displacement increases.

\bigskip

\begin{figure*}
  \centering
  \subfloat[]{\includegraphics[width=0.45\textwidth]{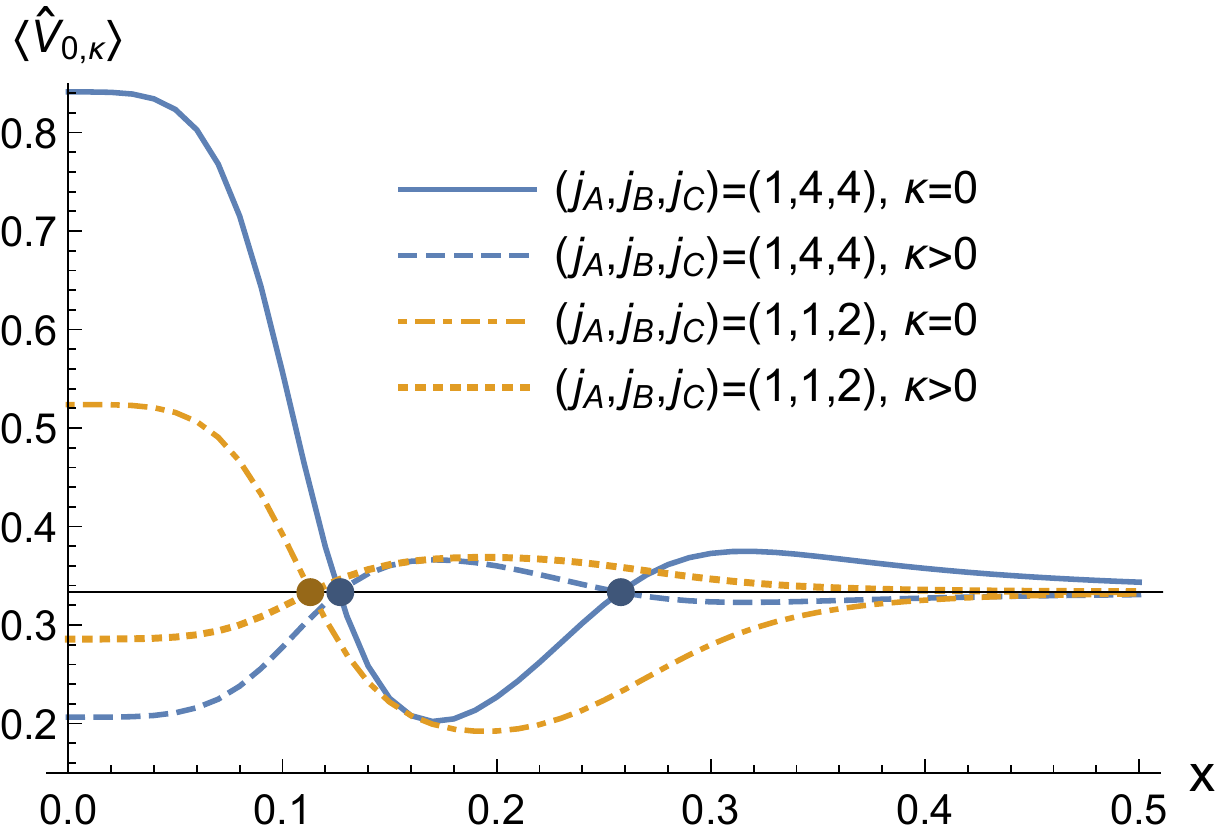}}
  \hfill
  \subfloat[]{\includegraphics[width=0.45\textwidth]{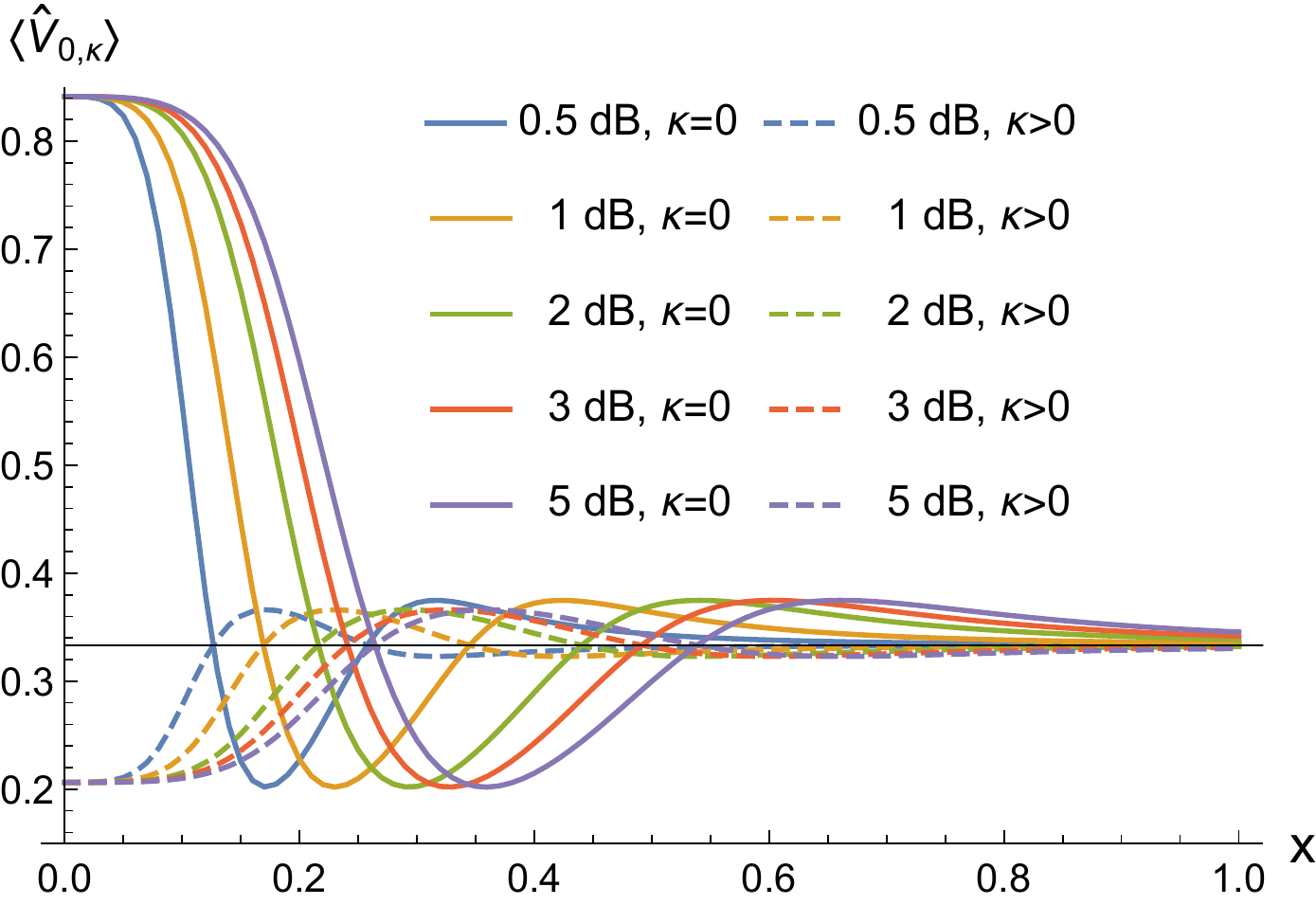}}
  \caption{%
    Evaluation of genuine multipartite entanglement of the $(5,5,5)$-mode $(2,1,1)$-photon state $\ket{\Psi_{2,1,1}(r,x)}$ given in Eq. \eqref{eq::ent_st_from_sq_input}.
    (a) The expectation value $\braket{\widehat{V}_{0,\kappa}}$ of $\ket{\Psi_{2,1,1}(r,x)}$ with a squeezing factor of $0.5$ dB.
    %a squeezing factor $r = 0.0576(0.5dB)$.
    The expectation value $\braket{\widehat{V}_{0,\kappa}}$ is evaluated in the $\widehat{\Lambda}_{j_{A}l}\otimes\widehat{\Lambda}_{j_{B}l}\otimes\widehat{\Lambda}_{j_{C}l}$-eigenbasis measurement settings with $(j_{A},j_{B},j_{C}) = (1,4,4)$ and $(j_{A},j_{B},j_{C}) = (1,1,2)$, respectively.
    (b) The expectation value $\braket{\widehat{V}_{0,\kappa}}$ of $\ket{\Psi_{2,1,1}(r,x)}$ is evaluated in the $\widehat{\Lambda}_{l}\otimes\widehat{\Lambda}_{4l}\otimes\widehat{\Lambda}_{4l}$-eigenbasis measurement settings for different squeezing.
  }%
  \label{fig::ENT_detection_SqSt}
\end{figure*}

\begin{figure*}
  \centering
  \subfloat[]{\includegraphics[width=0.88\textwidth]{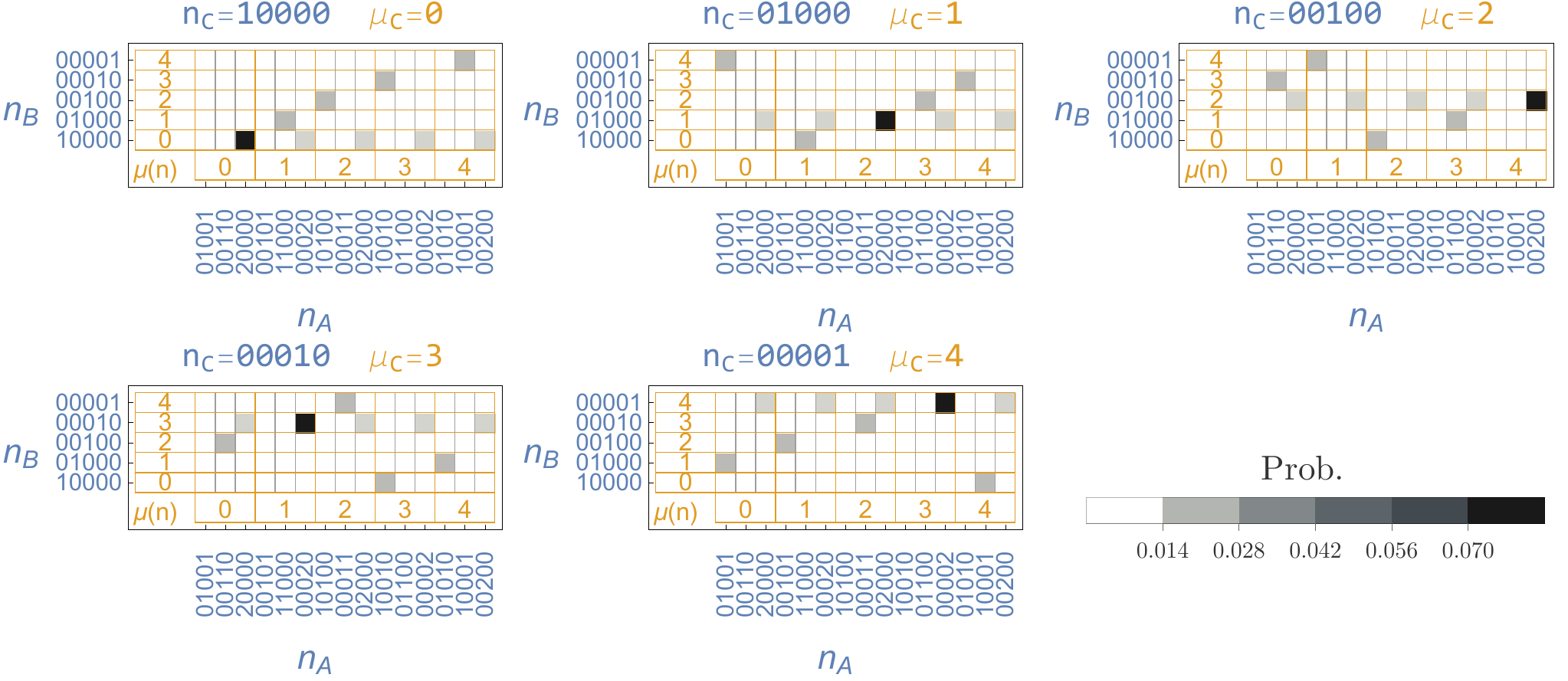}}
  \\
  \subfloat[]{\includegraphics[width=0.88\textwidth]{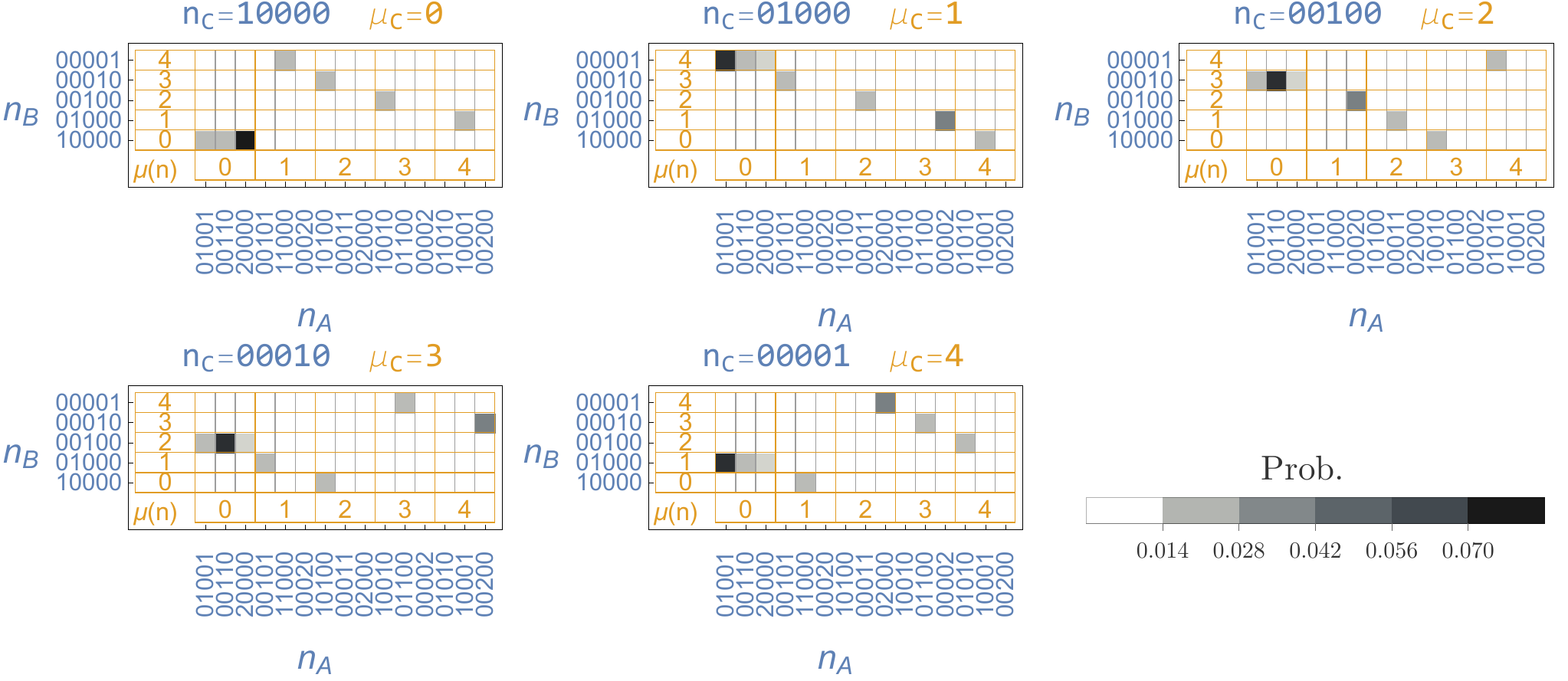}}
  \\
  \subfloat[]{\includegraphics[width=0.88\textwidth]{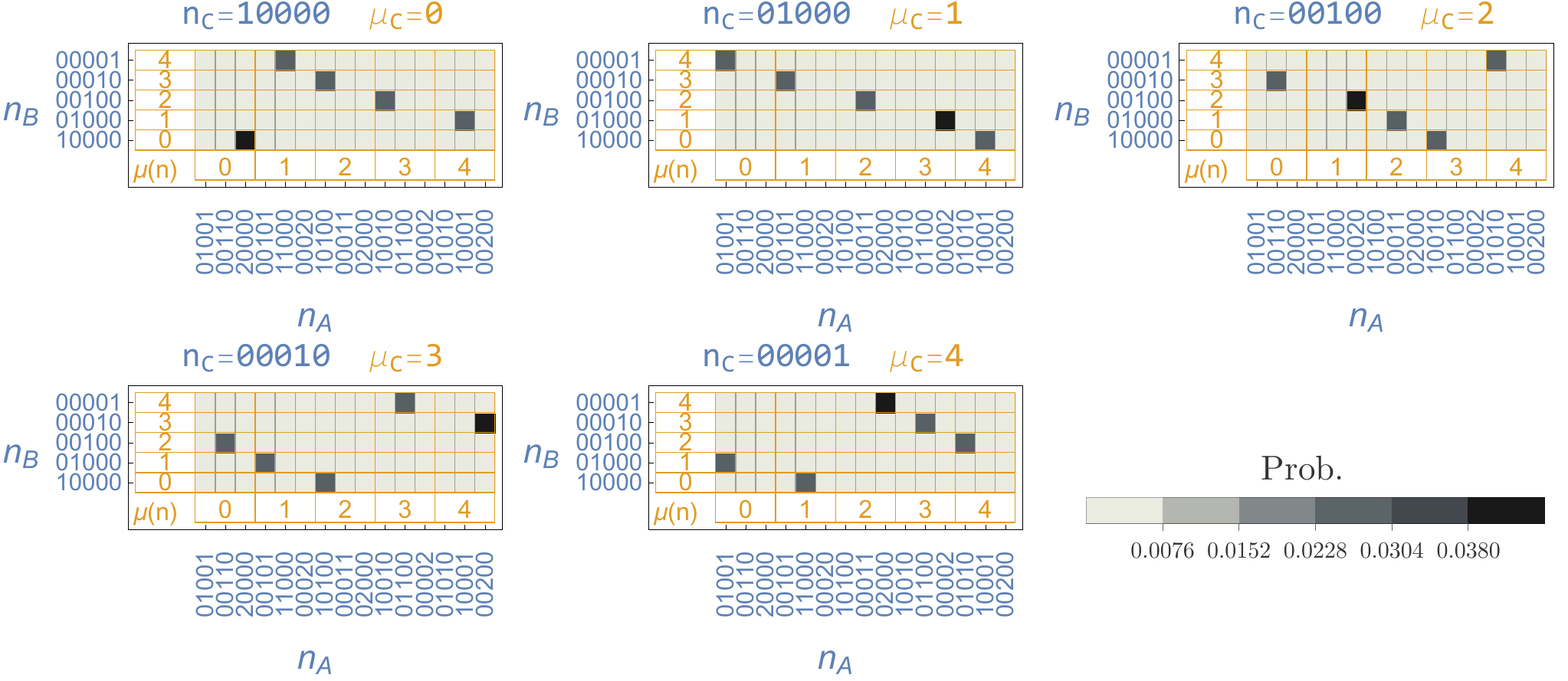}}
  \caption{Measurement statistics of the state $\ket{\Phi_{2,1,1}(r,x)}$ with a squeezing factor of $r=1.15\;(10dB)$ and a displacement of $x=0$.
  (a) Measurement in the computational basis.
  (b) The measurement in the $\widehat{\Lambda}_{0}\otimes\widehat{\Lambda}_{0}\otimes\widehat{\Lambda}_{0}$ eigenbasis.
  (c) The measurement in the $\widehat{\Lambda}_{j_{A}l}\otimes\widehat{\Lambda}_{j_{B}l}\otimes\widehat{\Lambda}_{j_{C}l}$ eigenbasis, where the HW indices are chosen as $(j_{A},j_{B},j_{C})=(1,4,4)$.}%
  \label{fig::M5_N211}
\end{figure*}

Here, we take a tripartite $(5,5,5)$-mode and $(2,1,1)$-photon system as an example.
The testing state $\ket{\Phi_{2,1,1}(r,x)}$ is a $(2,1,1)$-photon GM entangled state.
%For $x=0$, the post-selected state $\ket{\Phi_{2,1,1}(r,0)}$ does not depend on the squeezing factor $r$.
%For $x>0$ and $0<r<0.576 (5 dB)$, the squeezing factor will affect the form of the post-selected state $\ket{\Phi_{2,1,1}(r,x)}$.
%For a large squeezing $r>0.576 (5 dB)$, the form of the post-selected state $\ket{\Phi_{2,1,1}(r,x)}$ is asymptotical equal to $\ket{\Phi_{2,1,1}(\infty,x)}$.
To detect the GME, we first choose the HW indices as $(j_{A},j_{B},j_{C}) = (1,4,4)$.
The complementary measurement settings are implemented with the generalized Hadamard transformation $\widehat{H}_{l}\otimes\widehat{H}_{4l}\otimes\widehat{H}_{4l}$ with $l\in\mathbb{L} = \{0,...,M-1\}$. Together with the measurement in the computational basis, one will implement six measurements.
The measurement statistics of $\ket{\Phi_{2,1,1}(r,x)}$ is numerically simulated in Fig. \ref{fig::M5_N211} for $x=0$.

In Fig. \ref{fig::ENT_detection_SqSt} (a), we fix the squeezing factors as $r=0.058 (0.5 dB)$ and numerically evaluate the expectation values of the GME verifiers $\widehat{V}_{0,\kappa}$ for different displacements $x$ (the blue (dark gray) solid line for $\kappa=0$, and blue (dark gray) dashed line for $\kappa=1,...,4$).
%One can see that all expectation value $\braket{\widehat{V}_{0,\kappa}}$ sum up to $5/3$ as Eq. \eqref{eq::sum_of_GME_verifier} predicts.
The expectation value $\braket{\widehat{V}_{0,0}}$ has its maximum at $x=0$ and decreases as $x$ increases until a turning point, meanwhile, the expectation values $\braket{\widehat{V}_{0,\kappa=1,..,4}}$ have their minimum at $x=0$ and increases as $x$ increases until the same tuning point. There are two turning points before $\braket{\widehat{V}_{0,\kappa}}$ asymptotically approaches the bi-producible bound $1/3$.

Except for the two crossing points $x\approx 0.127$ and $x\approx0.257$, where $\braket{\widehat{V}_{0,\kappa=0}} = \braket{\widehat{V}_{0,\kappa>0}}=1/3$, there exists at least a $\braket{\widehat{V}_{0,\kappa}}$ greater than the bi-producible bound $1/3$.
This means that GME of $\ket{\Phi_{2,1,1}(r,x)}$ can always be detected for $x\neq0.127 \text{ or } 0.257$ with the measurement settings constructed in the $\widehat{\Lambda}_{l}\otimes\widehat{\Lambda}_{4l}\otimes\widehat{\Lambda}_{4l}$ eigenbasis.
For these two crossing points, one can verify the GME with the measurement settings constructed in other HW indices, e.g. $(j_{A},j_{B},j_{C}) = (1,1,2)$.
The expectation values $\braket{\widehat{V}_{0,\kappa}}$ evaluated in the $\widehat{\Lambda}_{l}\otimes\widehat{\Lambda}_{l}\otimes\widehat{\Lambda}_{2l}$-eigenbasis measurements are plotted with orange solid and dashed lines for $\kappa=0$ and $\kappa>0$, respectively.
One can see that the GME of $\ket{\Phi_{2,1,1}(r,x)}$ at $x=0.127$ and $x=0.257$ is verified by the GME verifier $\widehat{V}_{0,\kappa>0}$ in the $\widehat{\Lambda}_{l}\otimes\widehat{\Lambda}_{l}\otimes\widehat{\Lambda}_{2l}$-eigenbasis measurement settings.
%The numerical simulation in Fig. \ref{fig::ENT_detection_SqSt} (a) therefore agrees with Corollary \ref{coro::GME_detection_prime}.

In Fig. \ref{fig::ENT_detection_SqSt} (b), the expectation value $\braket{\widehat{V}_{0,\kappa}}$ evaluated in the $\widehat{\Lambda}_{l}\otimes\widehat{\Lambda}_{4l}\otimes\widehat{\Lambda}_{4l}$-eigenbasis measurements are plotted for different squeezing factors.
One can see that the squeezing factor changes the scale of the asymptotic behavior of $\braket{\widehat{V}_{0,\kappa}}$ versus the displacement.
As the squeezing factor increases, the value of $\braket{\widehat{V}_{0,\kappa}}(r,x)$ converges to a function of the displacement $x$, which is $r$-independent.

At the asymptotic limit, the photon statistics in the $(2,1,1)$-photon subspace behaves approximately as a multi-mode coherent state. This means no significant entanglement can be detected for a large displacement. Since the total entanglement of the CV system is conserved as the displacement increases, the diminishing of GME in the $(2,1,1)$-photon subspace indicates a transfer of GME from the $(2,1,1)$-photon subspace to other higher-photon-number subspaces.

Compared with the GME generated with single-photon inputs in Section \ref{sec::GME_gen_1ph}, the signature of the GME generated with CV inputs is not as significant as the one generated with discrete-variable sources. However, its scalability is much better than the scheme with single photons. In addition to these two aspects, photons losses play an important role in the GME generation with CV sources, as the postselection on local photon numbers can not exclude the photon-loss events from the statistics.

\subsection{Effect of photon losses}

In the GME generation and evaluation scheme in Fig. \ref{fig::GEM_gen_dect}, photons in each mode may be lost in the waveguide medium, at the beam splitters, or during the photon number resolving detection.
For the GME generation with single-photon sources in Section \ref{sec::GME_gen_1ph},
the effect of photon losses is excluded in the post-selected measurement statistics, since we post-select on an output total photon number which is equal to the input total photon number.
For the GME generation with CV sources, photon losses will change the measurement statistics and diminish the GME signature. In this case, one therefore has to take photon losses into account.

In linear optics networks, the mechanism of photon losses can be modeled as each path mode interferes with an external mode in the environment that is not accessible.
Such lossy interference can be modeled as a beam splitter with a reflection coefficient $\sqrt{\epsilon}$.
A photon has a probability $\epsilon$ of being reflected and lost in the environment (see Appendix \ref{sec::lossy_model} for detailed analysis).
For simplicity, we assume a uniform photon loss rate $\epsilon$ in each mode.
Employing the above lossy model, we obtain a lossy input state
\begin{align}
  & \widehat{\rho}_{N_{1},...,N_{P}}(r,x;\epsilon)
  \nonumber \\
  = &
  \sum_{\boldvec{\nu}_{1},...,\boldvec{\nu}_{P}}
  p_{\boldvec{\nu}_{i}|N_{i}}(\epsilon)
  \Projector{
    \Phi_{\boldvec{\nu}_{i}|N_{i}}(r,x)
    %\Phi_{N_{1},...,N_{P}}^{\boldvec{\nu}_{1}, ..., \boldvec{\nu}_{P}}
  },
\end{align}
where $p_{\boldvec{\nu}_{i}|N_{i}}(r,x;\epsilon)$ is the probability of losing $\boldvec{\nu}_{i}$ photons in the $i$-th local system, and the pure state component is
\begin{align}
  & \ket{\Phi_{\boldvec{\nu}_{i}|N_{i}}(r,x)}
  \nonumber \\
  = &
  \frac{1}{\sqrt{R_{\boldvec{\nu}_{i}|N_{i}}}}
  \sum_{\boldvec{n}_{i}:|\boldvec{n}_{i}|=N_{i}}
  \frac{
    h_{\boldvec{n}_{\text{tot}}+\boldvec{\nu}_{\text{tot}}}(2\zeta x)
  }{
    \sqrt{\prod_{i}\boldvec{n}_{i}!}
  }
  \ket{\boldvec{n}_{1},...,\boldvec{n}_{P}}.
\end{align}
with $R_{\boldvec{\nu}_{i}|N_{i}}$ being the normalization factor.
The expectation value of the lossy input state is therefore
\begin{align}
\label{eq::VGME_phLss}
  & \tr\left(\widehat{\rho}_{N_{1},...,N_{P}}^{ }(r,x;\epsilon)\;\widehat{V}_{0,\kappa}\right)
  \nonumber \\
  = &
  \sum_{\boldvec{\nu}_{1},...,\boldvec{\nu}_{P}}
  p_{\boldvec{\nu}_{i}|N_{i}}(r,x; \epsilon)
  \;
  \Braket{\Phi_{\boldvec{\nu}_{i}\vert N_{i}}(r,x)| \widehat{V}_{0,\kappa} |\Phi_{\boldvec{\nu}_{i}\vert N_{i}}(r,x)},
\end{align}
where $\braket{\widehat{V}_{0,\kappa}}$ of each component is
\begin{equation}
  \Braket{\Phi_{\boldvec{\nu}_{i}\vert N_{i}}(r,x)| \widehat{V}_{0,\kappa} |\Phi_{\boldvec{\nu}_{i}\vert N_{i}}(r,x)}
  =
  \frac{1+M|c_{0,\kappa}|^{2}}{M+1}.
\end{equation}
The expectation values $\braket{\widehat{V}_{0,\kappa}}$ of two states $\ket{\Phi_{\boldvec{\nu}_{i}|N_{i}}}$ and $\ket{\Phi_{\boldvec{\nu}'_{i}|N_{i}}}$ are identical if $\ket{\boldvec{\nu}_{\text{tot}}}$ and $\ket{\boldvec{\nu}'_{\text{tot}}}$ belong to the same $\widehat{X}$-irreducible class,
\begin{equation}
  \exists m\in\{0,...,M-1\},
  \;\;\text{ such that }
  \ket{\boldvec{\nu}_{\text{tot}}} = \widehat{X}^{m}\ket{\boldvec{\nu}'_{\text{tot}}}.
\end{equation}
In general, it holds that
\begin{equation}
  \left\{
    \begin{array}{ll}
      \sum_{\kappa}|c_{0,\kappa}|^{2}<1, & \hbox{for all} \ket{\boldvec{\nu}_{\text{tot}}}\neq\widehat{X}\ket{\boldvec{\nu}_{\text{tot}}}
      \\
      \sum_{\kappa}|c_{0,\kappa}|^{2}=1, & \hbox{for all} \ket{\boldvec{\nu}_{\text{tot}}} = \widehat{X}\ket{\boldvec{\nu}_{\text{tot}}}
    \end{array},
  \right.
\end{equation}
which means that the sum of the expectation value for different $\kappa$ under photon losses is in general smaller than $2M/(M+1)$ for non-zero photon losses,
\begin{equation}
  \sum_{\kappa}
  \tr\left(\widehat{\rho}_{N_{1},...,N_{P}}^{ }(r,x;\epsilon)\;\widehat{V}_{0,\kappa}\right)
  < \frac{2M}{M+1}
  \text{ for }
  \epsilon>0.
\end{equation}
As the photon loss rate increases, more non-$\widehat{X}^{\otimes P}$-symmetric components $\ket{\Phi_{\boldvec{\nu}_{i}|N_{i}}}$ with $\widehat{X}\ket{\boldvec{\nu}_{\text{tot}}}\neq \ket{\boldvec{\nu}_{\text{tot}}}$ are mixed in the state $\widehat{\rho}_{N_{1},...,N_{P}}$.
The significance of GME detection is therefore decreased by photon losses.

\bigskip

\begin{figure*}
  \centering
  \subfloat[]{\includegraphics[width=0.45\textwidth]{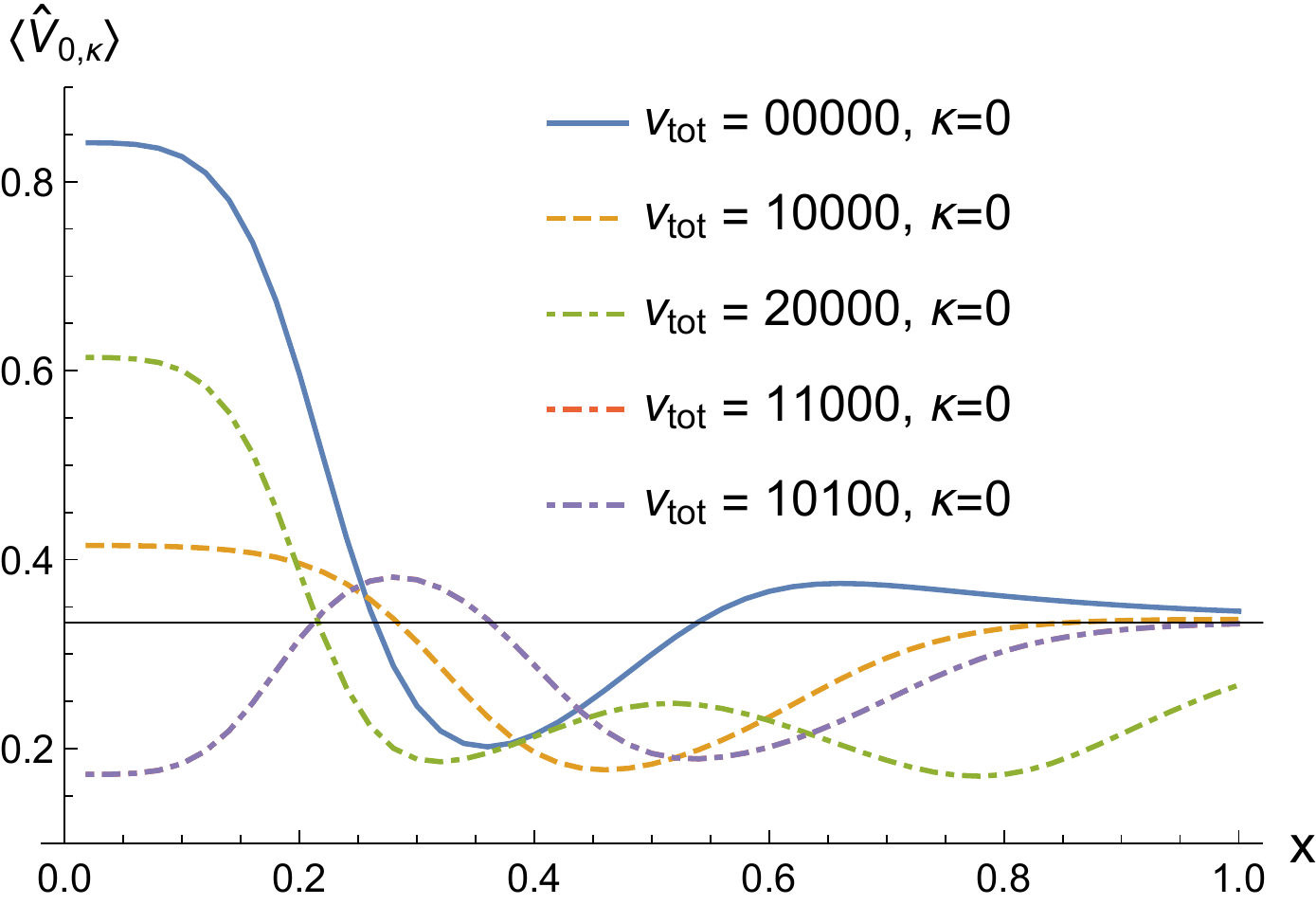}}
  \hfill
  \subfloat[]{\includegraphics[width=0.45\textwidth]{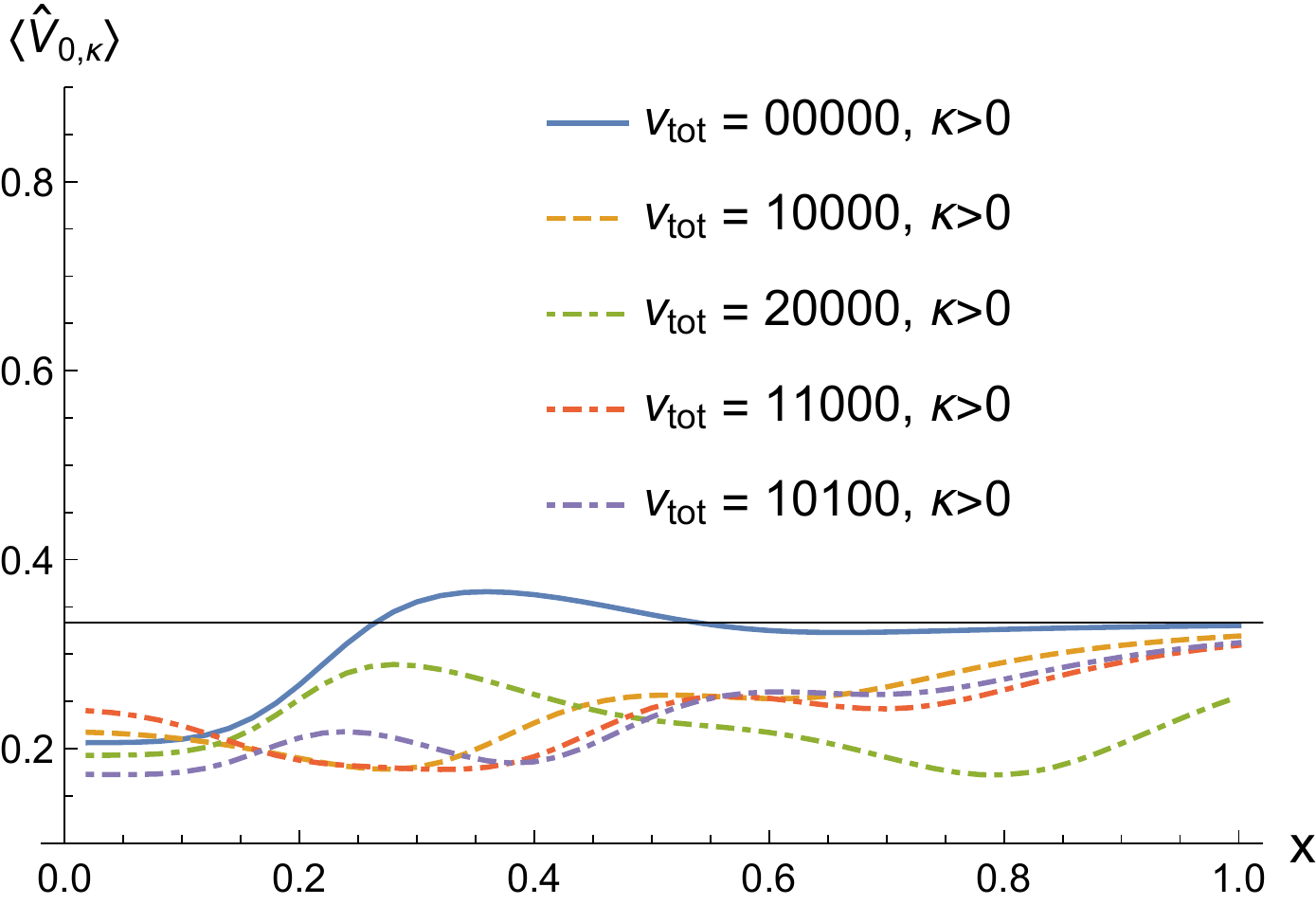}}
  \\
  \subfloat[]{\includegraphics[width=0.45\textwidth]{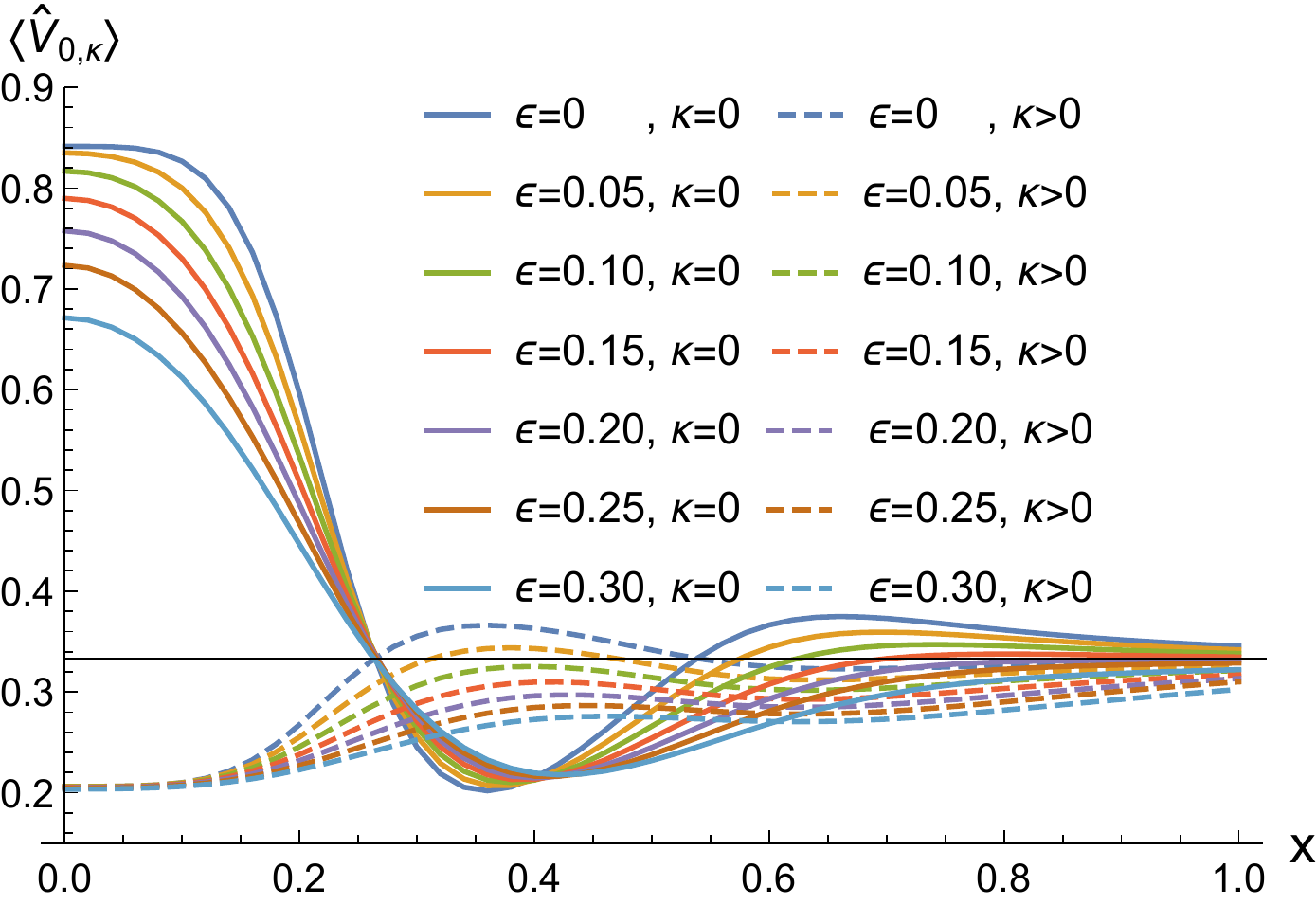}}
  \caption{%
  The expectation value $\braket{\widehat{V}_{0,\kappa}}$ with input squeezing factor $r = 0.576 (5 dB)$.
  (a) The expectation value $\braket{\widehat{V}_{0,\kappa}}$ of $\ket{\Phi_{\boldvec{\nu}_{i}|(2,1,1)}(r,x)}$ with $\kappa = 0$ and $\boldvec{\nu}_{tot} \in \{00000, 10000, 20000, 11000, 10100\}$.
  (b) The expectation value $\braket{\widehat{V}_{0,\kappa}}$ of $\ket{\Phi_{\boldvec{\nu}_{i}|(2,1,1)}(r,x)}$ with $\kappa > 0$ and $\boldvec{\nu}_{tot} \in \{00000, 10000, 20000, 11000, 10100\}$.
  (c) The expectation value $\braket{\widehat{V}_{0,\kappa}}$ of $\widehat{\rho}_{2,1,1}(r,x; \epsilon)$ for different loss rates.}
  \label{fig::phLss_effect}
\end{figure*}

For a small photon loss rate $\epsilon$, the expectation value in Eq. \eqref{eq::VGME_phLss} can be approximately evaluated with the sum over the photon-loss vector up to a small total photon loss.
In Fig. \ref{fig::phLss_effect}, we numerically evaluate the expectation value $\braket{\widehat{V}_{0,\kappa}}$ for $(5,5,5)$-mode $(2,1,1)$-photon GME generated with the input $x$-displaced $(5 \text{dB})$-squeezed state under the consideration of a uniform photon loss rate $\epsilon$ at each mode.
Fig. \ref{fig::phLss_effect} (a) and (b) show the $\braket{\widehat{V}_{0,\kappa}}$ of different components $\ket{\Phi_{\boldvec{\nu}_{i}|N_{i}}}$ with $\boldvec{\nu}_{\text{tot}} \in \{00000, 10000, 20000, 11000, 10100\}$ for $\kappa=0$ and $\kappa>0$, respectively.
For a small enough photon loss rate ($\epsilon\leq 0.25$), the expectation value $\braket{\widehat{V}_{0,\kappa=0}}$ of the mixed state $\widehat{\rho}_{N_{1},...,N_{P}}(r,x;\epsilon)$ for $\kappa=0$ and $\kappa>0$ can be approximately given by the convex combination of the values in Fig. \ref{fig::phLss_effect} (a) and (b), respectively.
%The approximation with $V_{\text{tot}}\le 2$ has no significant difference to the expectation value $\braket{\widehat{V}_{0,\kappa}}$ shown in Fig. \ref{fig::phLss_effect} (c), which is approximately evaluated with $V_{\text{tot}}\le 3$.
The sum of the expectation value for different $\kappa$ under photon losses is in general smaller than $5/3$,
\begin{equation}
  \sum_{\kappa}
  \tr\left(\widehat{\rho}_{N_{1},...,N_{P}}^{ }(r,x;\epsilon)\;\widehat{V}_{0,\kappa}\right)
  < 5/3
  \text{ for }
  \epsilon>0.
\end{equation}
The sum $\sum_{\kappa}\braket{\widehat{V}_{0,\kappa}}$ decreases as the photon loss rate increases.
For $\epsilon\ge0.2$ one can detect GME only for $x<0.265$.

%\subsection{Effect of photon losses}
%\begin{theorem}
%  Photon losses do not increase the GME signature?
%\end{theorem}
% Photon losses in input, processor, or detectors?
%\whatis{The photon losses in generation, interferometer, and detectors can be model as beam splitters align before photon being absorbed and the energy being transmitted into electron energy}
%
%Let $\eta$ be the probability of one photon being loss at an output mode.
%The state with one photon loss
  %%%%%%%%%%%%%%%%%%%%%%%%%%%%%%%%%%%%%%%%%%%%%%%%%%%%%%%%%%%%%%%%%%
  %%    Conclusion
\section{Discussion and Conclusion}
\label{sec::conclusion}
In this paper, we have derived a method for detecting genuine multipartite entanglement (GME) among multiphoton multimode linear optics networks in Theorem \ref{theorem::GME_detection}, which employs GME verifiers for a target GM-entangled state to reveal the signature for GME in experiments. The expectation value of a GME verifier exceeding its bi-producible bounds signifies the existence of GME. In general, the bi-producible bounds depend on the dimension of local mode-shifting irreducible subspaces, which can be determined adaptively to the measurement in the computational basis. The determination of these bi-producible bounds can be efficiently solved in classical computer without calculating permanents. For prime-number local modes, the upper bounds on the state verifiers within all local mode-shifting-irreducible subspaces are uniform. In this case, the upper bounds can be simplified as shown in Corollary \ref{coro::GME_detection_prime}.

A scheme for GME generation employing multimode splitters with indistinguishable single-photon or continuous-variable sources has been also proposed. Although the GME generation with single-photon sources can provide high entanglement signature, which is also robust to photon losses, the scalability of such GME generation requires a large number of single-photon sources with good indistinguishability. This obstructs the scalability of GME generation. On the other hand, although GME generation with continuous-variable sources provides a smaller GME signature, which is also diminished by photon losses, the generation scheme with continuous-variable sources is much more scalable than the one with single-photon sources.

The GME signature of $(2,1,1)$-photon $(5,5,5)$-mode GME generation with displaced squeezed vacuum inputs has been numerically investigated. Since the displacement operation on each input mode is equivalent to a local unitary, the overall GME of the continuous-variable system is conserved. However, the discrete-variable GME signature of the $(2,1,1)$-photon subspace diminishes as the displacement increases. One can obtain the most significant GME with a squeezed vacuum without displacement, while for a large displacement, one will not get a GME signature with a detectable contrast anymore. This implies a transfer of discrete-variable GME from lower-photon-number subspaces to higher-photon-number subspaces under the conservation of total continuous-variable entanglement.
%With the same detection setup for $(2,1,1)$-photon, one can also evaluate the GME in other fixed local-photon-number subspaces.

Our results provide a method to generate and evaluate GME in boson sampling systems with (discrete-variable) single-photon sources or (continuous-variable) squeezed-state sources. This theory allows us to reveal the discrete-variable GME in fixed local-photon-number subspaces of a continuous-variable system. It may help us understand the link between continuous-variable and discrete-variable entanglement in multiphoton linear optics networks. We may also expect possible applications of multipartite multi-rail linear optics networks for quantum conference key distribution based on GME distribution over quantum networks \cite{DasEtAlHorodecki2021-LmtOnQKD}. In addition, as GME is a result of photon indistinguishability in linear optics networks, it can be reversely exploited to characterize the input sources, which paves the way towards device-independent characterization of linear optics networks.

\acknowledgments
This work is supported by Ministry of Science and Technology, Taiwan, R.O.C. under Grant no. MOST 110-2112-M-032-005-MY3, 111-2923-M-032-002-MY5 and 111-2119-M-008-002.
\section{Appendix}

\subsection{Proof of Theorem \ref{theorem::GME_detection}}
\label{sec::proof}
According to \cite{WuMurao2020-CmplPropLONs}, the GME verifier $\widehat{V}_{k,\kappa}$ is diagonal with respect to $\widehat{X}$-irreducible subspaces,
\begin{equation}
  \widehat{V}_{k,\kappa}
  =
  \sum_{\mathbb{X}_{1},...,\mathbb{X}_{P}}
  \widehat{\pi}_{\mathbb{X}_{1},...,\mathbb{X}_{P}}
  \;\;\widehat{V}_{k,\kappa}\;\;
  \widehat{\pi}_{\mathbb{X}_{1},...,\mathbb{X}_{P}}
\end{equation}
where $\widehat{\pi}_{\mathbb{X}_{1},...,\mathbb{X}_{P}}$ is a projector onto the local $\widehat{X}$-irreducible subspaces $\bigotimes_{i}\spn(\mathbb{X}_{i})$, %$\mathbb{H}_{\mathbb{X}_{1}}\otimes\cdots\otimes\mathbb{H}_{\mathbb{X}_{P}}$,
\begin{equation}
  \widehat{\pi}_{\mathbb{X}_{1},...,\mathbb{X}_{P}}
  :=
  \sum_{\ket{\boldvec{n}_{i}}\in\mathbb{X}_{i}}
  \projector{\boldvec{n}_{1}, ..., \boldvec{n}_{P}}.
\end{equation}
The expectation value $\braket{\widehat{V}_{k,\kappa}}$ of a multiphoton state $\widehat{\rho}$ is therefore equal to the convex mixture of the expectation value $\braket{\widehat{V}_{k,\kappa}}$ of its projected component $\widehat{R}_{\mathbb{X}_{1},...,\mathbb{X}_{P}}(\rho)$,
\begin{equation}
  \tr\left(\widehat{V}_{k,\kappa} \rho \right)
  =
  \sum_{\mathbb{X}_{1},...,\mathbb{X}_{P}}
  p_{\mathbb{X}_{1},...,\mathbb{X}_{P}}(\rho)
  \tr\left(
    \widehat{V}_{k,\kappa}
    \widehat{R}_{\mathbb{X}_{1},...,\mathbb{X}_{P}}(\rho)
  \right),
\end{equation}
where $\widehat{R}_{\mathbb{X}_{1},...,\mathbb{X}_{P}}(\rho)$ is the state projected from $\widehat{\rho}$ onto the subspace $\bigotimes_{i}\spn(\mathbb{X}_{i})$, and
$p_{\mathbb{X}_{1},...,\mathbb{X}_{P}}$ is the probability of the projection,
\begin{equation}
  p_{\mathbb{X}_{1},...,\mathbb{X}_{P}}(\rho)
  =
  \tr\left(
    \widehat{\pi}_{\mathbb{X}_{1},...,\mathbb{X}_{P}}
    \;
    \widehat{\rho}
    \;
    \widehat{\pi}_{\mathbb{X}_{1},...,\mathbb{X}_{P}}
  \right)
\end{equation}
and
\begin{equation}
  \widehat{R}_{\mathbb{X}_{1},...,\mathbb{X}_{P}}(\rho)
  :=
  \frac{
    \widehat{\pi}_{\mathbb{X}_{1},...,\mathbb{X}_{P}}
    \;
    \widehat{\rho}
    \;
    \widehat{\pi}_{\mathbb{X}_{1},...,\mathbb{X}_{P}}
  }{p_{\mathbb{X}_{1},...,\mathbb{X}_{P}}(\rho)}
  .
\end{equation}
The upper bound $\beta$ on $\widehat{V}_{k,\kappa}$ for bi-producible states is therefore the convex extension of the upper bound $\beta_{\mathbb{X}_{1},...,\mathbb{X}_{P}}$ on $\widehat{V}_{k,\kappa}$ bi-producible states within each $\widehat{X}$-irreducible subspace $\bigotimes_{i}\spn(\mathbb{X}_{i})$,
\begin{align}
\label{eq::upper_bound_convex_roof}
  \beta(\rho)
  & :=
  \max_{\ket{\psi} \text{is bi-separble}}
  \braket{\psi|\widehat{V}_{k,\kappa}|\psi}
  \nonumber \\
  & =
  \sum_{\mathbb{X}_{1},..., \mathbb{X}_{P}}
  p_{\mathbb{X}_{1},...,\mathbb{X}_{P}}(\rho)
  \beta_{\mathbb{X}_{1},...,\mathbb{X}_{P}}
\end{align}

The upper bound on $\braket{\widehat{V}_{k,\kappa}}$ for a bi-separable pure state in an $\widehat{X}$-irreducible subspace $\bigotimes_{i}\spn(\mathbb{X}_{i})$ is
determined as follows.
Let $A|B$ be a bipartition of the $P$ local parties with $A\subset\{1, .., P\}$ and $B=\{1,...,M\}\setminus A$.
For a bi-separable pure state $\ket{\psi_{A},\psi_{B}}\in\,\bigotimes_{i}\spn(\mathbb{X}_{i})$,
In general, the expectation value $\braket{\widehat{V}_{k,\kappa}}$ is given as
\begin{equation}
  \braket{\widehat{V}_{k,\kappa}}
  =
  \frac{1}{1+|\mathbb{L}|}
  \sum_{k',\kappa}(\delta_{\kappa'}^{\kappa} + \sum_{l}\delta_{k'+\kappa'l}^{k+\kappa l})
  |c_{k',\kappa'}|^{2},
\end{equation}
where the probability amplitude  $|c_{k',\kappa'}|$ is equal to
\begin{equation}
  |c_{k',\kappa'}|^{2}
  =
  \sum_{\mu_{\boldvec{j}}(\mathbb{X}') = \kappa', \mathbb{X}'\subseteq\bigotimes_{i}\mathbb{X}_{i}}
  |\braket{\mathbb{E}_{k'}(\mathbb{X}')|\psi_{A},\psi_{B}}|^{2}.
\end{equation}
It holds that
\begin{equation}
  \delta_{\kappa'}^{\kappa} + \sum_{l}\delta_{k'+\kappa'l}^{k+\kappa l}
  \left\{
    \begin{array}{ll}
      \le 1 & \hbox{for $(k',\kappa')\neq(k,\kappa)$;} \\
      = 1 + |\mathbb{L}|, & \hbox{for $(k',\kappa')=(k,\kappa)$.}
    \end{array}
  \right.,
\end{equation}
which leads to the following inequality,
\begin{equation}
  \braket{\widehat{V}_{k,\kappa}}
  \le
  \frac{1+|\mathbb{L}| |c_{k,\kappa}|^{2}}{1+|\mathbb{L}|}.
\end{equation}
%The partial trace of a pure $(A|B)$ bi-separable state on $A$
%\begin{equation}
%  \ket{\varphi_{A}}\ket{\varphi_{B}}
%  =
%  \sum_{k',\kappa'} c_{k',\kappa'} \ket{\psi_{k',\kappa'}}
%\end{equation}
According to Eq. \eqref{eq::GM_ent_component}, an HW-symmetric state $\ket{\mathbb{E}_{k}(\mathbb{X}')}$ can be written as the Schmidt decomposition in the computational basis $\ket{\boldvec{n}_{j}}\in\mathbb{X}_{j}$
\begin{align}
  & \ket{\mathbb{E}_{k}(\mathbb{X})}
  %\nonumber \\
  = &
  \frac{1}{\sqrt{|\mathbb{X}_{j}|}}
  \sum_{\ket{\boldvec{n}_{j}}\in\mathbb{X}_{j}}
  \ket{\varphi_{1}(\boldvec{n}_{j})
  ,...,
  \boldvec{n}_{j}
  ,...,
  \varphi_{P}(\boldvec{n}_{j})}.
\end{align}
In this decomposition, the set of states $\{\ket{\varphi_{i}(\boldvec{n}_{j})}: \boldvec{n}_{j}\in\mathbb{X}_{j}\}$ is orthonormal.
It was shown in \cite{SpenglerHuberEtAlHiesmayr2012-EntWitViaMUB} that
\begin{equation}
  |\braket{\mathbb{E}_{k}(\mathbb{X}')|\psi_{A},\psi_{B}}|^{2}
  \le
  \frac{1}{|\mathbb{X}_{j}|}
\end{equation}
for all bi-separable $\ket{\psi_{A},\psi_{B}}$.
The upper bound for bi-producible states in $\bigotimes_{i}\spn(\mathbb{X}_{i})$ is therefore
\begin{equation}
  \beta_{\mathbb{X}_{1},...,\mathbb{X}_{P}}
  =
  \frac{1+|\mathbb{L}| / \min_{i}|\mathbb{X}_{i}|}{1+|\mathbb{L}|}.
\end{equation}
As a result of the convex roof extension in Eq. \eqref{eq::upper_bound_convex_roof}, we obtain the upper bound
\begin{equation}
  \beta(\rho)
  =
  \sum_{\boldvec{n}_{1},...,\boldvec{n}_{P}}
  \frac{
    \braket{\boldvec{n}_{1},...,\boldvec{n}_{P}|\widehat{\rho}|\boldvec{n}_{1},...,\boldvec{n}_{P}}
  }{
    \min_{i}|\mathbb{X}_{\boldvec{n}_{i}}|
  },
\end{equation}
which completes the proof.

\subsection{A lossy model in linear optics networks}
\label{sec::lossy_model}

\begin{figure*}
  \centering
  \subfloat[]{\includegraphics[width=0.33\textwidth]{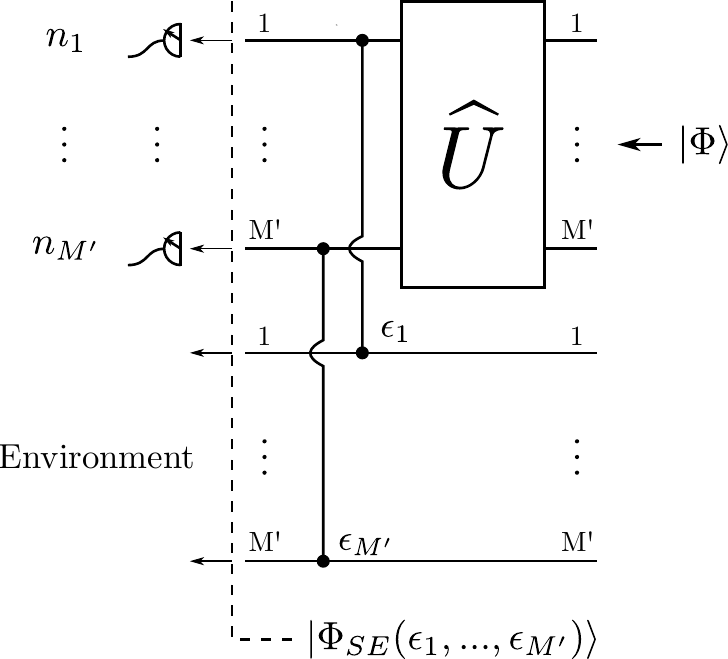}}
  \hfill
  \subfloat[]{\includegraphics[width=0.33\textwidth]{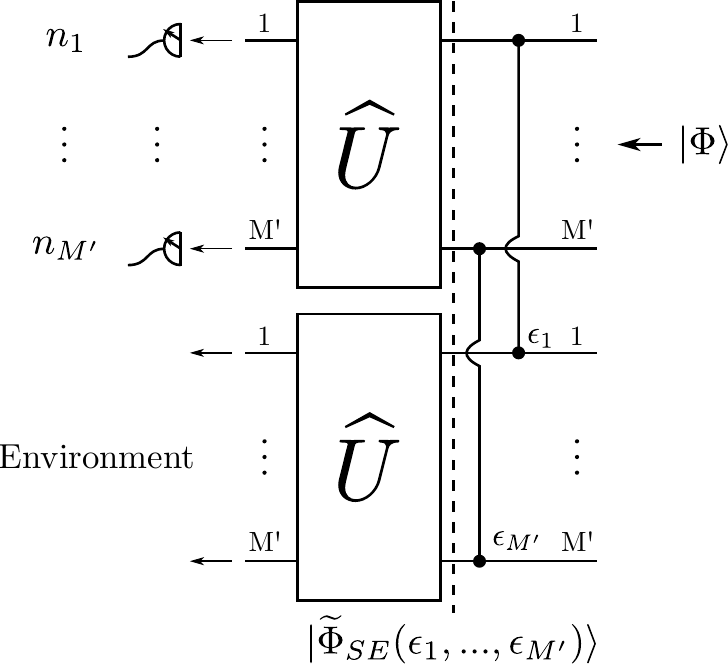}}
  \hfill
  \subfloat[]{\includegraphics[width=0.33\textwidth]{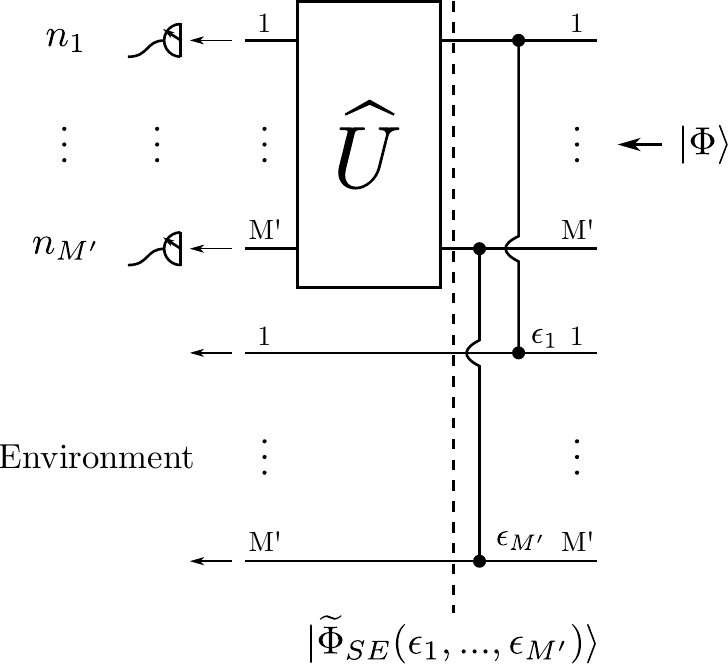}}
  \caption{
    (a) A lossy PNR measurement in LONs modeled by beam splitters interfering with the environment.
    (b) A circuit that equivalent to Fig. (a).
    (c) A circuit that equivalent to Fig. (b) after tracing out the environment.
  }%
  \label{fig::photon_losses}
\end{figure*}

As it is shown in Fig. \ref{fig::photon_losses} (a), the photon losses in PNR detectors of an $M'$-mode linear optics network can be modeled as $M'$ links connected to $M'$ external modes in the environment. Each link represents a beam splitter with a reflection coefficient $\sqrt{\epsilon_{m}}$.
%For an arbitrary unitary transformation on the $M'$-mode LON,
The circuit in Fig. \ref{fig::photon_losses} (a) is equivalent to the circuit in Fig. \ref{fig::photon_losses} (b),
\begin{equation}
\label{eq::loss_U_exhange}
  \prod_{m}\widehat{L}_{m}(\epsilon_{m}) \left(\widehat{U}_{\text{sys}}\otimes\id_{\text{env}}\right)
  =
  \left(\widehat{U}_{\text{sys}}\otimes\widehat{U}_{\text{env}}\right) \prod_{m}\widehat{L}_{m}(\epsilon_{m}).
\end{equation}
Here $\widehat{U}_{\text{sys}}$ and $\widehat{U}_{\text{env}}$ are the unitaries in the $M'$-mode system and environment, respectively.
The operator $\widehat{L}_{m}$ is the beam-splitter transformation which is responsible for the photon losses in the $m$-th mode,
\begin{equation}
  \widehat{L}_{m}\widehat{a}_{sys, m}^{\dagger}\widehat{L}_{m}^{\dagger}
  =
  \sqrt{1-\epsilon_{m}}\,\widehat{a}_{sys, m}^{\dagger}
  +
  \sqrt{\epsilon_{m}}\,\widehat{a}_{env, m}^{\dagger},
\end{equation}
where $\widehat{a}_{sys, m}^{\dagger}$ and $\widehat{a}_{env, m}^{\dagger}$ are the creation operator in the $m$-th mode of the system and environment, respectively.

The state at the output of Fig. \ref{fig::photon_losses} (a) is
\begin{equation}
  \ket{\Phi_{SE}(\epsilon_{1},...,\epsilon_{M'})} :=
  \prod_{m}\widehat{L}_{m}(\epsilon_{m})
  \left(\widehat{U}_{\text{sys}}\otimes\id_{\text{env}}\right)
  \ket{\Phi,\vac}.
%  =
%  \left(\widehat{U}_{\text{sys}}\otimes\widehat{U}_{\text{env}}\right)
%  \prod_{m}\widehat{L}_{m}(\epsilon_{m})
%  \ket{\Phi,\vac}.
\end{equation}
According to Eq. \eqref{eq::loss_U_exhange}, the photon losses can be shifted to the input side (see Fig. \ref{fig::photon_losses} (b))
\begin{equation}
  \ket{\Phi_{SE}}
  =
  \widehat{U}_{\text{sys}}\otimes\widehat{U}_{\text{env}}
  \ket{\widetilde{\Phi}_{SE}},
\end{equation}
where
\begin{equation}
  \ket{\widetilde{\Phi}_{SE}(\epsilon_{1},...,\epsilon_{M'})}
  :=
  \prod_{m}\widehat{L}_{m}(\epsilon_{m}) \ket{\Phi,\vac}.
\end{equation}
As a result, the output state in the LON system in Fig. \ref{fig::photon_losses} (a) is equivalent to the $\widehat{U}$ transformation of the input state with photon losses
\begin{equation}
  \tr_{\text{env}}\left(\projector{\Phi_{SE}}\right)
  =
  \widehat{U}
  \tr_{\text{env}}\left(\projector{\tilde{\Phi}_{SE}}\right)
  \widehat{U}.
\end{equation}
The right-hand side of this equation is equivalent to Fig. \ref{fig::photon_losses} (c), where the order of photon losses and LON transformation in Fig. \ref{fig::photon_losses} (a) is exchanged.
The effect of photon losses in a LON system can therefore be simply shifted to the input state $\ket{\widetilde{\Phi}_{SE}}$.
It leads to a mixed state input
\begin{align}
  \widehat{\rho}_{\Phi}(\epsilon_{1},...,\epsilon_{M'})
  & :=
  \tr_{\text{env}}\left(\projector{\widetilde{\Phi}_{SE}}\right)
  \nonumber \\
  & =
  \sum_{\boldvec{\nu}}
  p_{\boldvec{\nu}} \projector{\Phi_{\boldvec{\nu}}},
\end{align}
where $\boldvec{\nu}$ is the Fock vector of lost photons, $p_{\boldvec{\nu}}$ is the probability of losing $\boldvec{\nu}$ photons,
\begin{equation}
  p_{\boldvec{\nu}} = \tr\left( \braket{\boldvec{\nu}|\widetilde{\Phi}_{SE}}  \braket{\widetilde{\Phi}_{SE}|\boldvec{\nu}} \right),
\end{equation}
and
\begin{equation}
  \projector{\Phi_{\boldvec{\nu}}}
  =
  \frac{1}{p_{\boldvec{\nu}}}\braket{\boldvec{\nu}|\widetilde{\Phi}_{SE}}  \braket{\widetilde{\Phi}_{SE}|\boldvec{\nu}}.
\end{equation}
For a low photon loss rate, the lossy input state $\widehat{\rho}_{\Phi}(\epsilon_{1},...,\epsilon_{M'})$ can be
approximated through a cut-off up to a lost photon number $|\boldvec{\nu}|\le V_{\text{cutoff}}$,
\begin{equation}
  \widehat{\rho}_{\Phi}
  \approx
  \sum_{\boldvec{\nu}: |\boldvec{\nu}|\le V_{\text{cutoff}}|}
  p_{\boldvec{\nu}} \projector{\Phi_{\boldvec{\nu}}}.
\end{equation}
One can then numerically estimate the effects of photon losses through this approximation.

\newpage
%%                                                                  %%
%%                         Main text                                %%
%%%%%%%%%%%%%%%%%%%%%%%%%%%%%%%%%%%%%%%%%%%%%%%%%%%%%%%%%%%%%%%%%%%%%%

%%%%%%%%%%%%%%%%%%%%%%%%%%%%%%%%%%%%%%%%%%%%%%%%%%%%%%%%%%%%%%%%%%%%%%
%%                         Glossary                                 %%
%%                                                                  %%
\myprintglossary
%%                                                                  %%
%%                         Glossary                                 %%
%%%%%%%%%%%%%%%%%%%%%%%%%%%%%%%%%%%%%%%%%%%%%%%%%%%%%%%%%%%%%%%%%%%%%%

%%%%%%%%%%%%%%%%%%%%%%%%%%%%%%%%%%%%%%%%%%%%%%%%%%%%%%%%%%%%%%%%%%%%%%
%%                         Bibliography                             %%
%%                                                                  %%
\myprintbibliography
%%                                                                  %%
%%                         Bibliography                             %%
%%%%%%%%%%%%%%%%%%%%%%%%%%%%%%%%%%%%%%%%%%%%%%%%%%%%%%%%%%%%%%%%%%%%%%

\end{document}